\documentclass[prd,twocolumn,nopacs,floatfix,amsmath,nofootinbib,amssymb,preprintnumbers]{revtex4-2}
\usepackage{graphicx,subfigure,color,dcolumn,booktabs,diagbox}
\usepackage{txfonts}
\usepackage{slashed}
\usepackage{epstopdf}
\usepackage{multirow}
\usepackage{bm}
\usepackage{pifont}
\usepackage{adjustbox}
\usepackage{rotating}
\usepackage{verbatim}
\usepackage{graphicx,xcolor,dcolumn,booktabs}
\usepackage[colorlinks,
            citecolor=blue,
            anchorcolor=red,
            menucolor=red,
            linkcolor=red,
            filecolor=red,
            runcolor=red,
            urlcolor=blue,
            frenchlinks=true]{hyperref}
\begin{document}
\title{Coupled-channel study of $4S$-$3D$ mixing dynamics in $\psi(4220)$ and $\psi(4380)$}
\author{Zi-Long Man$^{1,2,3,4}$}
\author{Si-Qiang Luo$^{1,2,3,4}$}
\author{Zi-Yue Bai$^{1,2,3,4}$}
\author{Xiang Liu$^{1,2,3,4}$}
\email{xiangliu@lzu.edu.cn}

\affiliation{
$^1$School of Physical Science and Technology, Lanzhou University, Lanzhou 730000, China\\
$^2$Lanzhou Center for Theoretical Physics,
Key Laboratory of Theoretical Physics of Gansu Province,
Key Laboratory of Quantum Theory and Applications of MoE,
Gansu Provincial Research Center for Basic Disciplines of Quantum Physics, Lanzhou University, Lanzhou 730000, China\\
$^3$MoE Frontiers Science Center for Rare Isotopes, Lanzhou University, Lanzhou 730000, China\\
$^4$Research Center for Hadron and CSR Physics, Lanzhou University and Institute of Modern Physics of CAS, Lanzhou 730000, China}

\begin{abstract}
Among charmoniumlike $XYZ$ states, the $\psi(4220)$ and $\psi(4380)$ states have emerged as key candidates for exploring the charmonium spectrum. In this work, we propose a $4S$-$3D$ charmonium mixing scheme for the $\psi(4220)$ and $\psi(4380)$, induced by coupled-channel effects. By constructing a coupled-channel model, we identify the dynamical mechanism responsible for the large mixing angle observed in previous studies, which cannot be explained by conventional potential models alone. Our analysis reveals that the $DD_1$ channel significantly influences the lower state ($\psi(4220)$), while the $D^*D_1$ channel primarily affects the higher state ($\psi(4380)$). Furthermore, we investigate the two-body  Okubo-Zweig-Iizuka (OZI)-allowed strong decay behaviors of these states, providing insights into their total widths. This study not only supports the $4S$-$3D$ mixing scheme but also offers a deeper understanding of the role of coupled channels in shaping the charmonium spectrum above 4 GeV. Our results align with experimental observations and provide a framework for interpreting future data on charmonium states.
 
\end{abstract}

\maketitle

\section{Introduction}

The non-perturbative aspects of the strong interaction are closely linked to the study of hadron spectroscopy. Over the past two decades, a series of new hadronic states, including the charmoniumlike $XYZ$ states, has sparked extensive discussions about exploring the exotic hadronic zoo and constructing the conventional hadron family \cite{Liu:2013waa, Chen:2016qju, Chen:2016spr, Guo:2017jvc, Liu:2019zoy, Brambilla:2019esw, Chen:2022asf}. Among these observed charmoniumlike $XYZ$ states, the ones directly produced in $e^+e^-$ annihilation are referred to as the $Y$ states, which are distinguished by their masses. The challenge of understanding these $Y$ states has come to be known as the “$Y$ Problem”, as highlighted in the BESIII White Paper \cite{BESIII:2020nme}.

The first $Y$ state, the $Y(4260)$, was reported by the BaBar Collaboration in 2005, where an enhancement in the cross-section of $e^+e^-\to\pi^+\pi^-J/\psi$ was observed. This structure, with a mass around 4.26 GeV and a width of approximately 90 MeV, was identified as a vector state \cite{BaBar:2005hhc}. Later, the CLEO \cite{CLEO:2006tct} and Belle \cite{Belle:2007dxy} Collaborations confirmed the existence of the $Y(4260)$. Several theoretical explanations have been proposed for its nature, including the hybrid state \cite{Zhu:2005hp,Kou:2005gt}, the compact tetraquark state \cite{Maiani:2005pe,Ebert:2008kb,Chen:2015dig,Chen:2010ze,Zhang:2010mw}, and the molecular state \cite{Yuan:2005dr,Cleven:2013mka,Close:2010wq,Li:2013yla,Lu:2017yhl}. However, these resonance assignments face significant experimental challenges. Notably, the $Y(4260)$ structure is absent in open-charm decay channels \cite{Belle:2006hvs,Belle:2007qxm,Belle:2007xvy,Belle:2009dus}, and there is a dip structure observed around 4.26 GeV in $R$ value measurements \cite{CLEO:2008ojp,BES:2009ejh,PLUTO:1976jbe,DASP:1978dns,Siegrist:1981zp,BES:1999wbx}.

In 2011, the Lanzhou Group proposed a nonresonant explanation for the $Y(4260)$ structure, attributing it to the interference between two well-established charmonia, $\psi(4160)$ and $\psi(4415)$, along with background contributions from the continuum. They showed that this interference could effectively reproduce the asymmetric $Y(4260)$ structure. This nonresonant explanation naturally accounts for the absence of evidence for the $Y(4260)$ in open-charm decay channels and $R$ value data \cite{Chen:2010nv}. Furthermore, this interference mechanism is considered universal, as it can also explain the $Y(4360)$ structure, observed in the $e^+e^-\to \psi(3686)\pi^+\pi^-$ process \cite{BaBar:2006ait}, through a similar interference effect \cite{Chen:2011kc,Chen:2015bft}. 

Clearly, this is not the final chapter of the story. In 2014, the Lanzhou Group observed a similarity in the mass gaps between the $J/\psi$ and $\Upsilon$ families \cite{He:2014xna}. Based on the mass gap between the $\Upsilon(4S)$ and $\Upsilon(3S)$, and the charmonium state $\psi(4040)\equiv \psi(3S)$, they predicted the mass of the $\psi(4S)$ to be $4264$ MeV \cite{He:2014xna}. This differs from the $\psi(4415)$ assignment as the $\psi(4S)$ in the quenched potential model \cite{Eichten:1979ms}, but is consistent with predictions from some unquenched potential models \cite{Dong:1994zj,Ding:1995he,Li:2009zu}, where the screening potential accounts for unquenched effects. Notably, they suggested that the predicted $\psi(4S)$, with a mass around 4264 MeV, would have a narrow width due to the node effect \cite{He:2014xna}. This prediction is yet to be tested experimentally.

Meanwhile, experimentalists identified evidence for a narrow structure with a mass of $4216 \pm 18$ MeV and a width of $39 \pm 32$ MeV by fitting the experimental data of the $e^+e^- \to \pi^+ \pi^- h_c$ cross sections \cite{Yuan:2013uta}. This narrow structure is considered a strong candidate for the predicted $\psi(4S)$ \cite{He:2014xna}. In 2015, the BESIII Collaboration reported a narrow resonance in the $e^+e^- \to \omega \chi_{c0}$ process, with a mass of $4230 \pm 8 \pm 6$ MeV and a width of $38 \pm 12 \pm 2$ MeV. The Lanzhou Group proposed the $\psi(4S)$ to explain the enhancement observed in the $e^+e^- \to \omega \chi_{c0}$ process \cite{Chen:2014sra}. Later, a combined fit to experimental data from $e^+e^- \to \psi(2S) \pi^+ \pi^-$ \cite{Belle:2014wyt}, $h_c \pi^+ \pi^-$ \cite{BESIII:2013ouc}, and $\omega \chi_{c0}$ \cite{Chen:2014sra} further refined the resonance parameters of the narrow structure around 4.2 GeV, which is consistent with the predicted $\psi(4S)$.

Experimental precision must match theoretical precision, often leading to surprising results. In 2017, the BESIII Collaboration revealed that the $Y(4260)$ structure consists of two substructures, the $Y(4220)$ and $Y(4320)$, based on more precise data from $e^+e^- \to J/\psi \pi^+ \pi^-$ \cite{BESIII:2016bnd}. The first substructure, the $Y(4220)$, has a mass of $4222.0 \pm 3.1 \pm 1.4$ MeV and a width of $44.1 \pm 4.3 \pm 2.0$ MeV. Additionally, the narrow $Y(4220)$ was also observed in the $e^+e^- \to h_c \pi^+ \pi^-$ \cite{BESIII:2016adj} and $e^+e^- \to \psi(3686) \pi^+ \pi^-$ \cite{BESIII:2017tqk} processes. These combined theoretical and experimental efforts suggest that the narrow $Y(4220)$ is a strong candidate for the predicted $\psi(4S)$.
 
When carrying out a quantitative study of charmonium spectroscopy by a quenched potential model by introducing a screening potential, as did in Ref. \cite{He:2014xna}, the calculated mass of the $\psi(4S)$ of charmonium is about 50 MeV larger than the measured value of the $Y(4220)$ \cite{Wang:2019mhs}. 

\begin{figure}[!htbp]
	\centering
	\begin{minipage}[b]{0.48\textwidth}
		\centering
		\includegraphics[width=1\textwidth]{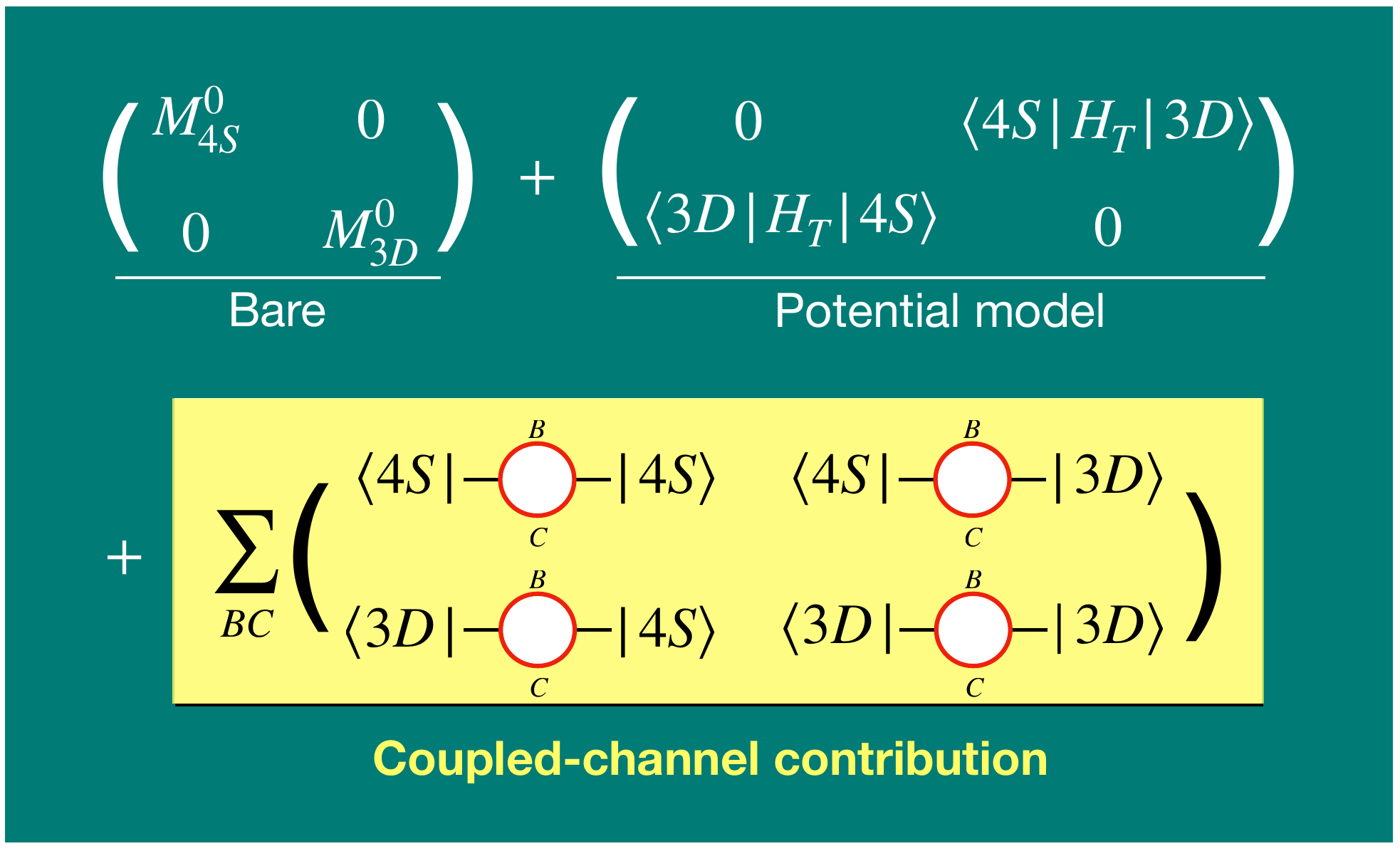}
	\end{minipage}
	\caption{Schematic diagram illustrating the $4S$-$3D$ mixing in the potential model and the coupled-channel mechanism. For the coupled-channel contribution, hadronic loops involving $B$ and $C$ mesons are considered in the calculation, which will be presented in Sec. \ref{sec3}. \label{mass}}
\end{figure}

Obviously, this discrepancy cannot be overlooked. Given that the $2S$-$1D$ mixing scheme has been successfully applied to study the spectroscopic behaviors of the two low-lying charmonia $\psi(3686)$ and $\psi(3770)$, it is reasonable to believe that a similar $S$-$D$ mixing scheme could be relevant in understanding the nature of the $Y(4220)$. In this context, the Lanzhou Group introduced a $4S$-$3D$ mixing scheme for the $Y(4220)$ \cite{Wang:2019mhs}, showing that the mass of the mixed state can be lowered to match that of the $Y(4220)$ when the mixing angle is in the range of $\pm(30^\circ \sim 36^\circ)$. Additionally, a second mixed state, $\psi(4380)$, was predicted. In fact, the $Y(4220)$ serves as a scaling point for constructing higher charmonium states above 4.2 GeV. Subsequently, a $5S$-$4D$ mixing scheme for the $\psi(4415)$ was proposed, predicting its partner, $\psi(4500)$. This approach results in an unquenched mass spectrum of vector charmonia above 4 GeV, comprising states such as $\psi(4040)$, $\psi(4160)$, $\psi(4220)$, $\psi(4380)$, $\psi(4415)$, and $\psi(4500)$. These states could be observed \cite{Wang:2019mhs,Wang:2022jxj,Wang:2023zxj,Peng:2024xui} in cross-section data from processes such as $e^+e^- \to \psi(3686)\pi^+\pi^-$ \cite{BESIII:2017tqk}, $e^+e^- \to \pi^+ D^0 D^{*-}$ \cite{BESIII:2018iea}, $e^+e^- \to J/\psi \eta$ \cite{BESIII:2023tll}, and $e^+e^- \to J/\psi K^+K^-$ \cite{BESIII:2022joj}.

While the $4S$-$3D$ mixing scheme has been successfully applied to explain the $Y(4220)$ phenomenologically \cite{Wang:2019mhs}, it is essential to identify a mechanism that can explain the large mixing angle observed in the $4S$-$3D$ charmonium mixing scheme. This angle is sufficiently large and cannot be reproduced by the tensor term in the potential model \cite{Godfrey:1985xj} (see the second term in Fig. \ref{mass}). Therefore, the main motivation for this investigation is to uncover the underlying dynamical mechanism responsible for this large mixing angle, which will further elucidate the $4S$-$3D$ mixing scheme for the $\psi(4220)$.

In this work, we propose a coupled-channel induced charmonium mixing scheme for the $\psi(4220)$ as illustrated in the schematic diagram in Fig. \ref{mass}. By constructing a coupled-channel model for the $\psi(4220)$ and its partner, the $\psi(4380)$, we successfully reproduce the large mixing angle observed in Ref. \cite{Wang:2019mhs}. However, in contrast to previous studies that employed an unquenched potential model with a screening potential \cite{Wang:2019mhs}, our approach demonstrates how the coupled channels influence the masses of the bare states. Our calculations reveal that the $DD_1$ channel plays a significant role in the lower state ($\psi(4220)$), while the nearby $D^*D_1$ channel contributes primarily to the higher state ($\psi(4380)$). Additionally, we examine the two-body  Okubo-Zweig-Iizuka (OZI)-allowed strong decay behaviors of the physical states $\psi(4220)$ and $\psi(4380)$, which help determine the total width of these states.

The paper is organized as follows: In Sec. \ref{sec2}, we introduce a potential model to study the bare masses of $\psi(4S)$ and $\psi(3D)$. Then, we employ a comprehensive coupled-channel model to discusses $4S-3D$ mixing scheme in \ref{sec3}. The final section provides a brief summary.

%%%%%%%%%%%%%%%%%%%%%%%%%%%%%%%%%%%%%%%%%%%%%%%%%%%%%
\section{Bare mass of the discussed charmonia from a potential and the corresponding mixing scheme induced by the tensor term}\label{sec2}
%%%%%%%%%%%%%%%%%%%%%%%%%%%%%%%%%%%%%%%%%%%%%%%%%%%%%

We begin by calculating the bare masses of the discussed charmonia using a potential model, as these results serve as essential inputs for the subsequent analysis within the coupled-channel framework. Various versions of potential models have been employed to describe the hadron mass spectrum, most of which are derived from the Cornell potential \cite{Godfrey:1985xj, Barnes:2005pb,Godfrey:2015dva, Godfrey:2015dia, Godfrey:2014fga, Godfrey:2016nwn}. Among these, the Godfrey-Isgur (GI) model stands out as it meets the precision requirements necessary for the study of the current hadron mass spectrum.

The GI model, a typical quenched potential model, is a semi-relativistic framework with the Hamiltonian given by
\begin{align}\label{GI}
H_{0}=(\boldsymbol{p}^2+{m_c}^2)^{\frac12}+(\boldsymbol{p}^2 +{m_{\bar{c}}}^2)^{\frac12}+\tilde{V}_{\rm{ eff}}(\boldsymbol{p},\boldsymbol{r}),
\end{align}
where \( m_{c} \) and \( m_{\bar{c}} \) represent the masses of the \( c \) and \( \bar{c} \) quarks, respectively. The effective potential \( \tilde{V}_{\rm eff}(\boldsymbol{p}, \boldsymbol{r}) \) incorporates a short-range one-gluon-exchange term \( \gamma^{\mu}\otimes\gamma_{\mu} \) and a long-range confinement term \( 1\otimes1 \). In the nonrelativistic limit, \( \tilde{V}_{\rm eff}(\boldsymbol{p},\boldsymbol{r}) \) reduces to:
\begin{align}\label{2.2}
V_{\rm eff}(r)=H^{\rm conf}+H^{\rm hyp}+H^{\rm so},
\end{align}
where the spin-independent term \( H^{\rm conf} \) combines the linear confinement potential and the Coulomb potential:
\begin{align}\label{2.3}
H^{\rm conf}=br-\frac{4\alpha_s(r)}{3r}+c.
\end{align}

The color-hyperfine interaction \( H^{\text {hyp }} \) in Eq. \eqref{2.2} consists of two components:
\begin{equation}\begin{aligned}
    H^{\text {hyp }}= H_S+H_T,
    \end{aligned}\end{equation}
where the spin-spin interaction \( H_S \) is given by
\begin{equation}\begin{aligned}
   H_{S}= & \frac{32\pi\alpha_S(r)}{9m_c m_{\bar{c}}} \boldsymbol{S}_c \cdot \boldsymbol{S}_{\bar{c}} \delta^3(\boldsymbol{r}),
    \end{aligned}\end{equation}
and the tensor term \( H_T \) is expressed as
\begin{equation}\begin{aligned}\label{tensor term}
H_T=\frac{4\alpha_s}{3m_c m_{\bar{c}}r^3}
\left(\frac{3(\boldsymbol{S}_c \cdot \boldsymbol{r}) (\boldsymbol{S}_{\bar{c}} \cdot \boldsymbol{r})}{r^2}-\boldsymbol{S}_{c} \cdot \boldsymbol{S}_{\bar{c}}\right),
\end{aligned}\end{equation}
which may result in $S$-$D$ mixing of charmonium.
Here, \( \boldsymbol{S}_{c} \) and \( \boldsymbol{S}_{\bar{c}} \) denote the spins of the \( c \)-quark and \( \bar{c} \)-quark, respectively. The spin-orbit interaction \( H^{\rm so} \) in Eq. \eqref{2.2} is further divided into two terms
\begin{align}
H^{\rm so}=H^{\rm so(cm)}+H^{\rm so(tp)},
\end{align}
where the color-magnetic term \( H^{\rm so(cm)} \) and the Thomas precession term \( H^{\rm so(tp)} \) are given by
\begin{align}\label{2.5}
H^{\rm so(cm)}=\frac{4\alpha_{s}(r)}{3r^{3}}\left(\frac{ \boldsymbol{S}_{c}}{m^2_{c}}+\frac{\boldsymbol{S}_{\bar{c}}}{m^2_{{\bar{c}}}}+\frac{
\boldsymbol{S}_{c}+\boldsymbol{S}_{\bar{c}}}{m_{c}m_{\bar{c}}}\right) \cdot\boldsymbol{L},
\end{align}
and
\begin{align}
H^{\rm so(tp)}=-\frac{1}{2 r} \frac{\partial H^{\rm conf}}{\partial r}\left(\frac{ \boldsymbol{S}_{c}}{m_{c}^{2}}+\frac{ \boldsymbol{S}_{\bar{c}}}{m_{{\bar{c}}}^{2}}\right) \cdot\boldsymbol{L},
\end{align}
respectively. Here, \( \boldsymbol{L} \) represents the relative orbital angular momentum between the quark and antiquark. A detailed description of these equations can be found in Ref. \cite{Godfrey:1985xj}.

The parameters of the GI model are fitted from the masses of charmonia. In Ref. \cite{Duan:2020tsx}, the parameter values were determined, and the masses of the low-lying charmonium states were well reproduced. Therefore, we adopt the same parameters in this study, which are listed in Table \ref{parameter}. For further details on the fitted parameters, refer to Ref. \cite{Duan:2020tsx}.

\begin{table}[!htbp]
\renewcommand\arraystretch{1.5}
	\caption{The parameters of the GI model.}\centering
	\begin{tabular*}{1.0\columnwidth}{@{\extracolsep{\fill}}cccc@{}}
		\toprule[1.00pt]
		\toprule[1.00pt] \label{parameter}
		Parameters& Values     & Parameters &  Values\\
		 \midrule[0.75pt]
		$m_q$     & 0.220 GeV  &  $b$         & 0.175 $\text{GeV}^2$\\
            $m_c$     & 1.628 GeV  & $c$          & $-0.245$ GeV            \\
		$m_s$     & 0.419 GeV  &  $\epsilon_{\text{cont}}$& $-0.103$ \\
            $s$       & 0.821 GeV  &        $\epsilon_{\text{so(v)}}$  &  $-0.279$          \\
	    $\sigma_0$& 2.33 GeV  &        $\epsilon_{\text{so(s)}}$  &  $ -0.3 $           \\
	    $\alpha_s$& 0.6         &        $\epsilon_{\text{tens}}$&  $ -0.114$       \\
	    $\Lambda$ & 0.2 GeV      &\\
	    \bottomrule[1.00pt]
	    \bottomrule[1.00pt] 
\end{tabular*}
\end{table}

With the above preparation, we may utilize the GI model to give the bare masses of the involved charmonia $\psi(4S)$ and $\psi(3D)$, which are $4433.0$ MeV and $4491.3$ MeV, respectively. The obtained mass of $\psi(4S)$ is $211$ MeV higher than that of the $\psi(4220)$, but close to that of the $\psi(4415)$. This is why the charmonium state $\psi(4415)$ is identified as a $\psi(4S)$ state within the quenched potential model \cite{Eichten:1979ms}. However, with the advent of high-precision hadron spectroscopy in the unquenched model, we find that this scenario changes, as demonstrated in Ref. \cite{Wang:2019mhs}, where the $\psi(4S)$ state is lowered to $4274$ MeV , closer to the observed $\psi(4220)$. In the next section, a detailed coupled-channel analysis further confirms this, offering a different perspective.

When introducing the $4S$-$3D$ mixing scheme \cite{Wang:2019mhs}, the following relation holds:
\begin{gather}
	\begin{pmatrix}
		|\psi^\prime_{4S-3D}\rangle \\ |\psi^{\prime\prime}_{4S-3D}\rangle
	\end{pmatrix}=
	\begin{pmatrix}
		\cos\theta & \sin\theta \\ -\sin\theta & \cos\theta
	\end{pmatrix}
	\begin{pmatrix}
		| 4 ^3S_1 \rangle \\ | 3 ^3D_1 \rangle
	\end{pmatrix},
\end{gather}
where $\theta$ represents the mixing angle. The lower state $|\psi^\prime_{4S-3D}\rangle$ corresponds to the $\psi(4220)$, while the higher state $|\psi^{\prime\prime}_{4S-3D}\rangle$ can be identified as the $\psi(4380)$ \cite{Wang:2019mhs}. 
Using the masses of $\psi(4S)$ and $\psi(3D)$ as inputs, the mixing angle can be determined to be $\pm(30^\circ\sim36^\circ)$. This approach is phenomenological in nature. 

Starting from the GI model, the tensor term can contribute to the mixing of $S$- and $D$-wave charmonium states. A similar situation arises in the case of the deuteron, where the bound state properties are closely related to the $S$-$D$ mixing induced by the tensor term in the effective potential that describes the proton-neutron interaction \cite{Rarita:1941zza}. 
In the following, we examine the contribution of the tensor term in Eq. \eqref{tensor term} to the mixing of $S$- and $D$-wave charmonium states by solving the equation:
\begin{gather}\label{tensor}
	\begin{pmatrix}
	M_S^0	& \langle \psi_S|H_{\text{T}} |\psi_D\rangle\\
    \langle \psi_D|H_{\text{T}} |\psi_S\rangle& M_D^0
	\end{pmatrix}
    \begin{pmatrix}
		C_S \\ C_D \\ 
	\end{pmatrix}= M \begin{pmatrix}
		C_S \\ C_D \\
	\end{pmatrix}.
\end{gather}
The diagonal terms, $M_S^0$ and $M_D^0$, represent the bare masses of $\psi(4S)$ and $\psi(3D)$, respectively. According to our calculations, the $4S$-$3D$ mixing angle induced by the tensor term is only $0.5^\circ$, which is insufficient to account for the large mixing reported in Ref.~\cite{Wang:2019mhs}. This mixing angle has negligible impact on both the masses and decay properties of $\psi(4S)$ and $\psi(3D)$. This is why we need to find an alternative source for the $4S$-$3D$ mixing scheme, which will be the main focus of the following section.

\section{$4S$-$3D$ mixing scheme induced by the coupled-channel mechanism}\label{sec3}

Among the newly observed hadronic states, a universal phenomenon known as the ``low mass puzzle'' has been identified in the $X(3872)$ \cite{Belle:2003nnu}, $D_{s0}(2317)$ \cite{BaBar:2003oey}, $D_{s1}(2460)$ \cite{CLEO:2003ggt}, and $\Lambda_c(2940)$ \cite{BaBar:2006itc}. Specifically, the measured masses of these states are consistently lower than the masses predicted by quenched models when these states are treated as conventional hadrons (i.e., $q\bar{q}$ mesons or $qqq$ baryons). To address this low mass puzzle, a direct approach has been to consider the possibility of exotic hadronic configurations for these states. This approach has opened a new window into the study of hadron spectroscopy over the past few decades \cite{,Chen:2016qju,Guo:2017jvc,Liu:2019zoy,Chen:2022asf,Liu:2024uxn}.

In fact, the emergence of this low mass puzzle has highlighted the limitations of the quenched model, which was once the cornerstone of hadron spectroscopy. The influence of this type of model has persisted to the present day, but the observed discrepancies have prompted a reevaluation of its applicability.

When unquenched effects are taken into account, the low mass puzzle observed in the $X(3872)$, $D_{s0}(2317)$, $D_{s1}(2460)$, and $\Lambda_c(2940)$ can be significantly alleviated \cite{vanBeveren:2003kd,vanBeveren:2003jv,Kalashnikova:2005ui,Li:2009zu,Luo:2019qkm,Man:2024mvl}. This demonstrates that unquenched effects cannot be ignored, and in fact, should be emphasized in modern hadron spectroscopy. It is for this reason that the current era of hadron spectroscopy is increasingly defined by the unquenched model, marking a significant shift in our understanding of hadronic states.

In the study of hadron spectroscopy, there are different approaches to account for the unquenched effect. As demonstrated in previous studies \cite{Wang:2019mhs,Wang:2018rjg,Pan:2024xec}, one method involves introducing a screening potential within the framework of the potential model. Although this is a phenomenological approach, it effectively captures the realistic unquenched effect, as evidenced by recent applications in the study of higher-lying charmonium states \cite{,Wang:2018rjg,Wang:2020prx,Pan:2024xec}. Another approach is to perform a comprehensive coupled-channel analysis, which explicitly incorporates the coupling between the bare state and its allowed hadronic channels \cite{Lu:2016mbb,Fu:2018yxq}. The latter method is employed in the present work. It is worth noting that the equivalence of these two treatments has been approximately established in Refs. \cite{Li:2009ad,Duan:2021alw}.

Next, we perform a coupled-channel analysis for the $4 ^3S_1$ and $3 ^3D_1$ charmonium states and their mixing. This analysis is based on the following expression
\begin{widetext}
\begin{gather}
	\begin{pmatrix}
		M^0_S                                  & \langle \psi_S|H_{\text{T}} |\psi_D\rangle            &\sum_{BC}\int \langle \psi_S|H_I| BC,\mathbf{P} \rangle d^3\mathbf{P} \\
\langle \psi_D|H_{\text{T}} |\psi_S\rangle &M^0_D  
           &\sum_{BC}\int \langle \psi_D|H_I|BC,\mathbf{P} \rangle d^3\mathbf{P} \\
           \sum_{BC}\int \langle BC,\mathbf{P}|H_I|\psi_S\rangle d^3\mathbf{P}  
           &\sum_{BC}\int \langle BC,\mathbf{P}|H_I|\psi_D\rangle d^3\mathbf{P} &E_{BC}\\
	\end{pmatrix}
	\begin{pmatrix}
		C_S \\ C_D \\  C_{BC}
	\end{pmatrix}= M \begin{pmatrix}
		C_S \\ C_D \\ C_{BC}
	\end{pmatrix}.
\end{gather}
Here, $E_{BC}=\sqrt{M_B^2+\mathbf{P}^2}+\sqrt{M_C^2+\mathbf{P}^2}$ represents the energy of the meson continuum states, assuming that the interaction between $B$ and $C$ is negligible. The off-diagonal term $H_I$ describes the interactions between the meson-antimeson continuum states and the bare charmonium states. 

The mixing angle $\theta$ can be determined by solving the reduced coupled-channel equation
\begin{gather}\label{mix-couple}
	\begin{pmatrix}
		M_S^0+\Delta M_{S}(M)&
       \langle \psi_S|H_{\text{T}} |\psi_D\rangle+\Delta M_{SD}(M)\\
		\langle \psi_D|H_{\text{T}} |\psi_S\rangle+\Delta M_{SD}(M)&
        M_D^0+\Delta M_{D}(M)
	\end{pmatrix}
    \begin{pmatrix}
		C_S \\ C_D \\ 
	\end{pmatrix}= M \begin{pmatrix}
		C_S \\ C_D \\
	\end{pmatrix}.
\end{gather}
The forms of $\Delta M_{S}(M)$ and $\Delta M_{D}(M)$ are given by
\begin{eqnarray}\label{PiBC}
	\Delta M_{S(D)}(M)
	&=&\text{Re}\sum_{BC}\int_0^\infty\frac{\mid\langle \psi_{S(D)}| H_I| BC,P\rangle\mid^2P^2dP}{M-E_B-E_C},\nonumber\\
\end{eqnarray}
and $\Delta M_{SD}(M)$ is expressed as
\begin{equation}
	\Delta M_{SD}(M)= \text{Re} \sum_{BC}\int_0^\infty\frac{\langle \psi_S| H_I| BC,P\rangle\langle BC,P\mid H_I\mid \psi_D\rangle}{(M-E_B-E_C)}P^2dP.
\end{equation}
In principle, all possible open-charm meson loops should be included in the self-energy function. However, a challenge arises in calculating the infinite number of hadron loops in $\Delta M_{S(D)}(M)$ and $\Delta M_{SD}(M)$. This issue was addressed in Ref. \cite{Pennington:2007xr}, where the authors proposed the once-subtracted dispersion relation, which effectively limits the number of loops and resolves this problem. Here, we choose same number of hadronic loops to calculate mass shifts of mixing charmomium states, and these allowed loops with mass thresholds below the bare masses of $\psi(4S)$ and $\psi(3D)$.  

We now employ the once-subtracted method to rewrite $\Delta M_{S(D)}(M)$ and $\Delta M_{SD}(M)$ as
\begin{eqnarray}\label{oncePiBC}
	\Delta M_{S(D)}(M)
	&=&\text{Re} \sum_{BC}\int_0^\infty\frac{(M_{J/\psi}-M)|\langle \psi_{{S(D)}}^0| H_I|BC,P\rangle|^2P^2 dP}
	{(M-E_B-E_C)(M_{J/\psi}-E_B-E_C)},
\end{eqnarray}
and
\begin{equation}
	\Delta M_{SD}(M)= \text{Re} \sum_{BC}\int_0^\infty\frac{(M_{J/\psi}-M)\langle \psi_S\mid H_I\mid BC,P\rangle\langle BC,P\mid H_I\mid \psi_D\rangle}{(M-E_B-E_C)(M_{J/\psi}-E_B-E_C)}P^2dP,
\end{equation}
respectively. Here, $M_{J/\psi}$ denotes the subtraction point, which we take as the mass of the $J/\psi$ meson in the charmonium system. Studies using this method have successfully reproduced reasonable meson masses, as demonstrated in Refs. \cite{Duan:2020tsx,Duan:2021alw,Ni:2023lvx,Deng:2023mza}.

\end{widetext}

To describe the coupling between the bare state and the hadronic channel composed of $B$ and $C$, we express the matrix element as $\langle\psi| H_I|BC\rangle=\mathcal{M}_{JL} (A\rightarrow B+C)$. To calculate the amplitude $\mathcal{M}_{JL} (A\rightarrow B+C)$, we employ the quark pair creation (QPC) model \cite{LeYaouanc:1972vsx,LeYaouanc:1973ldf}. The corresponding Hamiltonian is given by
\begin{eqnarray}
	H_I&=&-3\gamma\sum_{m} \left<1m1-m\mid00\right>\int \limits d\mathbf{p}_3d\mathbf{p}_4\delta^3(\mathbf{p}_3+\mathbf{p}_4)\nonumber\\
	&&\times{\cal Y}_{1}^{m}\left(\frac{\mathbf{p}_3-\mathbf{p}_4}{2}\right)\chi_{1-m}^{34}\varphi_{0}^{34}\omega_{0}^{34}b_{3}^{\dagger}(\mathbf{p}_3)d_{4}^{\dagger}(\mathbf{p}_4),
\end{eqnarray}
where the dimensionless parameter $\gamma$ represents the strength of the $q\bar{q}$ pair creation from the vacuum. In our calculations, we adopt $\gamma=0.44$ as the strength parameter for the quark pair creation. This value is chosen because it reproduces the decay widths well of the two $P$-wave charmonium states, $\psi(3915) \equiv \psi(2 ^3P_0)$ and $\psi(3930) \equiv \psi(2 ^3P_2)$. If a strange quark pair is created, the effective strength is adjusted to $\gamma_s=\frac{m_n}{m_s}\gamma$, where $m_{n}$ ($n=u,d$) and $m_s$ denote the constituent masses of the nonstrange and strange quarks, respectively. The momenta of the created quark $q$ and antiquark $\bar{q}$ are denoted by $\mathbf{p}_3$ and $\mathbf{p}_4$, respectively. The solid harmonic function is given by ${\cal Y}_{1}^{m}(\mathbf{p})=|\mathbf{p}| Y_1^m(\theta,\phi)$, and $\chi_{1-m}^{34}$, $\varphi_{0}^{34}$, and $\omega_{0}^{34}$ represent the spin, flavor, and color wave functions of the created $q\bar{q}$ pair, respectively.

Using the amplitude $\mathcal{M}_{JL} (A\rightarrow B+C)$, the OZI-allowed two-body strong decay width for the charmonium state can be calculated as
\begin{eqnarray}\label{Im}
	\Gamma_\text{total}=\sum_{BC} \frac{2\pi PE_BE_C}{M}\sum_{JL}\mid\mathcal{M}_{JL} (A\rightarrow B+C)\mid^2,
\end{eqnarray}
where $P$ is the momentum of the final-state mesons, and $M$ is the mass of the physical state under consideration. In our calculations, the spatial wave functions of the involved mesons serve as crucial inputs. These wave functions are obtained by solving a potential model, as described in Sec. \ref{sec2}. Specifically, we adopt the following form for depicting the obtained numerical spatial wave function
\begin{eqnarray}\label{true wave function}
	\Psi_{n L M_{L}}(\mathbf{P}) &= \sum\limits_{n_1}^{n_{\text{max}}} C_n R_{nL}(P) Y_{LM_{L}}(\Omega_\mathbf{P}),
\end{eqnarray}
where $R_{nL}(P)$ represents the radial wave function in momentum space, and $n_{\text{max}}$ is the maximum number of basis states. To ensure an accurate representation of the numerical spatial wave functions, we set $n_{\text{max}}=20$.
This approach provides the necessary framework to determine the OZI-allowed two-body strong decay widths of charmonium states.

\subsection{The properties of $\psi(4S)$ and $\psi(3D)$ within coupled-channel analysis}

 \begin{figure}[!htbp]
	\centering
	\begin{minipage}[b]{0.48\textwidth}
		\centering
		\includegraphics[width=1\textwidth]{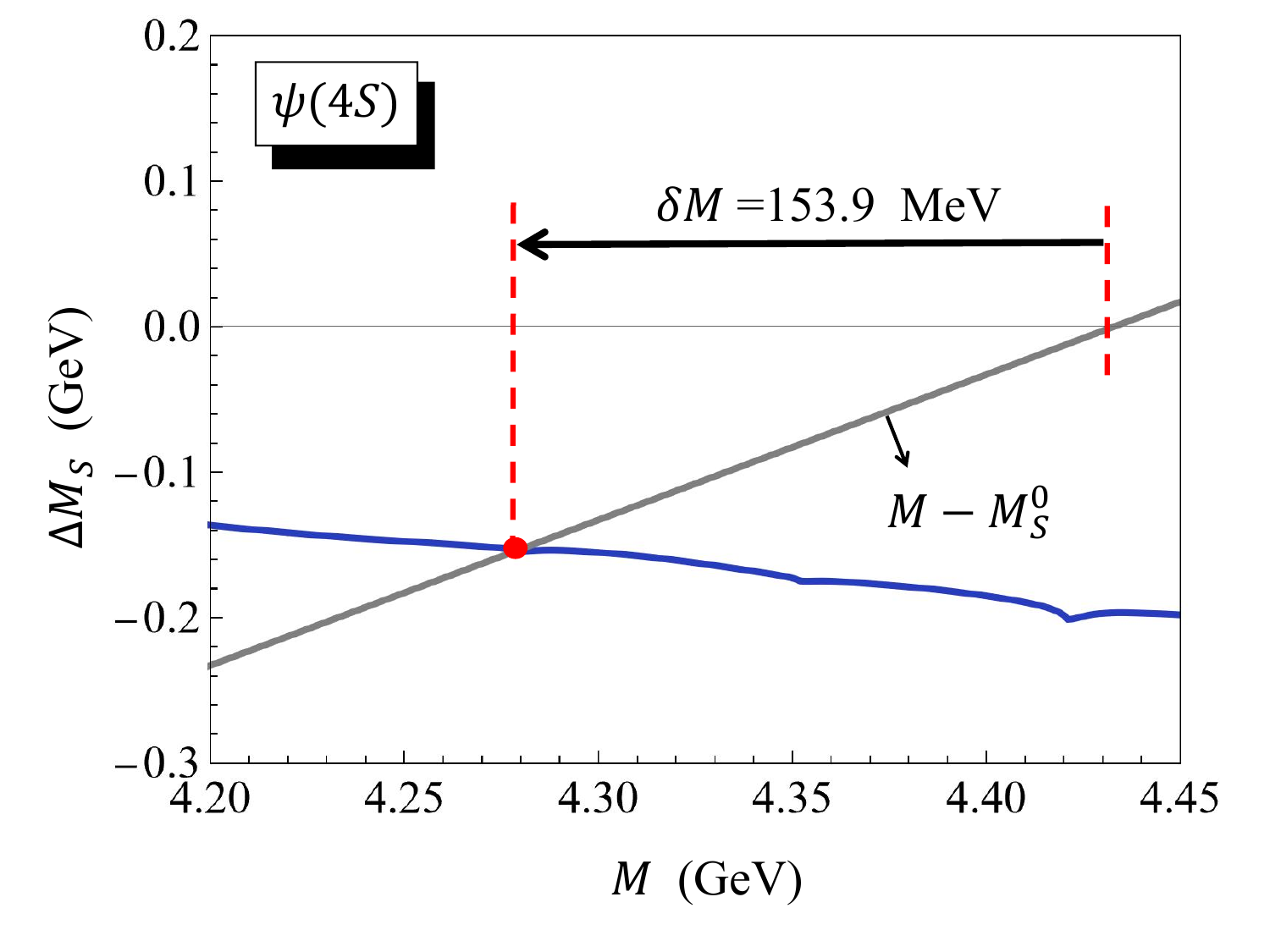}
        
	\end{minipage}
    \begin{minipage}[b]{0.48\textwidth}
		\centering
		\includegraphics[width=1\textwidth]{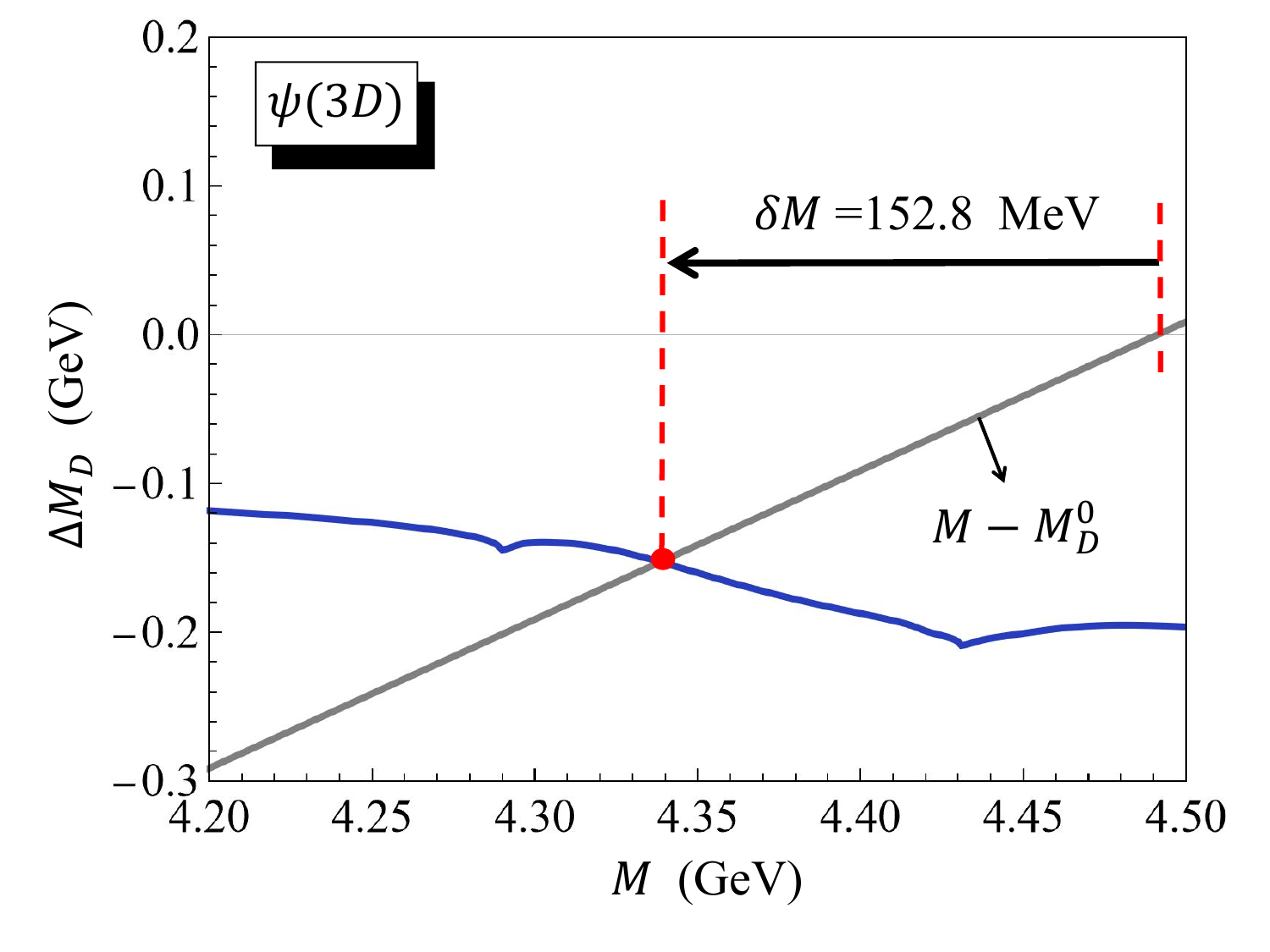}  
	\end{minipage}
	\caption{The $M$ dependence of function $M_0$ and $\Delta M(M)$ for $\psi(4S)$ and $\psi(3D)$, respectively.  There are three (two)  discontinuities in the line shape of $\psi(4S)$ ($\psi(3D)$) corresponding to the $DD_1(2430)^0$, $D^*D_1(2430)^0$ and $D^*D_0(2300)$ ($DD_1(2420)$ and $D^*D_1(2420)$) channels, respectively. }\label{4S}
\end{figure}

\begin{table*}[!htbp]
\renewcommand\arraystretch{1.5}
\caption{The mass shift $\Delta M_i$ from various coupled channels, the corresponding ratio $\Delta M_i/\sum_i \Delta M_i$, the partial decay widths, and branching ratios for $\psi(4S)$ and  $\psi(3D)$  within the coupled-channel framework. In the last two rows, we provide the average value from PDG \cite{ParticleDataGroup:2024cfk} for $\psi(4220)$ and the calculated resonance parameters for $\psi(4S)$. {The masses, mass shifts, and decay widths are  in units of MeV. }
 }\label{psi4S and psi3D}
\begin{tabular}{ccccccccc@{}}
\toprule[1.00pt]
\toprule[1.00pt]
\label{mass3D}
Channels &$\Delta{M}_i$&$\Delta{M}_i/\sum_i \Delta{M}_i$&$\Gamma_{i}$&$\Gamma_{i}/\sum_i \Gamma_{i}$ &$\Delta{M}_i$&$\Delta{M}_i/\sum_i \Delta{M}_i$&$\Gamma_{i}$&$\Gamma_{i}/\sum_i \Gamma_{i}$\\ \hline 
$DD$                                       & $-$5.8     & 3.8\%  & \
0.3        & 0.6\%  & $-$10.7    & 7.0\%  & 1.8        & 7.8\% \\
$DD^*$                                     & $-$12.3    & 8.0\%  & \
9.7        & 18.3\% & $-$3.1     & 2.0\%  & 3.4        & 14.4\%\\
$D_sD_s$                                   & $-$1.0     & 0.7\%  & \
0.2        & 0.3\%  & $-$1.1     & 0.7\%  & 0.4        & 1.9\% \\
$D^*D^*$                                   & $-$35.1    & 22.8\% & \
41.0       & 77.8\% & $-$36.9    & 24.2\% & 8.0        & 34.0\%\\
$D_sD_s^*$                                 & $-$4.4     & 2.8\%  & \
0.4        & 0.8\%  & $-$1.3     & 0.9\%  & 0.9        & 3.6\% \\
$DD_0^*(2300)$                             & $-$        & $-$   & $-$ \
       & $-$   & $-$        & $-$   & $-$        & $-$  \\
$D_s^*D_s^*$                               & $-$7.1     & 4.6\%  & \
1.1        & 2.1\%  & $-$10.1    & 6.6\%  & 0.7        & 3.1\% \\
$DD_1(2430)^0$                             & $-$16.1    & 10.4\% & \
$-$        & $-$   & $-$16.4    & 10.7\% & 2.4        & 10.2\%\\
$DD_1(2420)$                               & $-$8.7     & 5.7\%  & \
$-$        & $-$   & $-$14.5    & 9.5\%  & 5.7        & 24.0\%\\
$D_sD_{s0}(2317)$                          & $-$        & $-$   & $-$ \
       & $-$   & $-$        & $-$   & $-$        & $-$  \\
$DD_2^*(2460)$                             & $-$8.5     & 5.5\%  & \
$-$        & $-$   & $-$5.1     & 3.3\%  & 0.2        & 0.9\% \\
$D^*D_0^*(2300)$                           & $-$12.1    & 7.8\%  & \
$-$        & $-$   & $-$8.4     & 5.5\%  & $-$        & $-$  \\
$DD_0(2550)^0$                             & $-$2.2     & 1.4\%  & \
$-$        & $-$   & $-$6.2     & 4.1\%  & $-$        & $-$  \\
$D^*D_1(2430)^0$                           & $-$22.8    & 14.8\% & \
$-$        & $-$   & $-$21.4    & 14.0\% & $-$        & $-$  \\
$D_sD_{s1}(2460)$                          & $-$2.3     & 1.5\%  & \
$-$        & $-$   & $-$3.3     & 2.2\%  & $-$        & $-$  \\
$D_s^*D_{s0}(2317)$                        & $-$2.2     & 1.4\%  & \
$-$        & $-$   & $-$1.1     & 0.7\%  & $-$        & $-$  \\
$D^*D_1(2420)$                             & $-$13.2    & 8.6\%  & \
$-$        & $-$   & $-$13.0    & 8.5\%  & $-$        & $-$  \\
Total & $-$153.9 & 100\%  & 52.7     & 100\%  &$-$152.8   & 100\%  & 23.6       & 100\%  \\ 
    & $M=4279.1$ & & &  & $M=4338.5$ &  &  & \\\hline
     & $M_{\psi(4220)}=4222.1\pm2.3$ \cite{ParticleDataGroup:2024cfk}&   & $\Gamma_{\psi(4220)}=49\pm7$ \cite{ParticleDataGroup:2024cfk}&  &  & & &   \\
\bottomrule[1.00pt]
\bottomrule[1.00pt]
\end{tabular}
\end{table*}

{Before discussing the $4S$-$3D$ mixing scheme induced by the coupled-channel effect, it is essential to quantitatively investigate the impact of the coupled channels on the bare charmonium states $\psi(4S)$ and $\psi(3D)$. This analysis will help us better understand the fundamental features of the mass shifts for $\psi(4S)$ and $\psi(3D)$. }

Using the parameters outlined above, the numerical results for the masses, mass shifts, and decay widths of the $\psi(4S)$ and $\psi(3D)$ states are presented in Table \ref{psi4S and psi3D}. The mass shifts $\Delta M$ and the corresponding function $M_0-M$ as a function of $M$ for the $\psi(4S)$ and $\psi(3D)$ states are shown in Fig. \ref{4S}.

To quantitatively study the coupled-channel corrections to the bare $\psi(4S)$ and $\psi(3D)$ states, Eq. \eqref{mix-couple} can be decomposed into two separate equations, one for $\psi(4S)$ and the other for $\psi(3D)$. For $\psi(4S)$, the coupled-channel equation is expressed as:
\begin{eqnarray}\label{coupled-S}
	M - M_S^0 - \Delta M_S(M) = 0.
\end{eqnarray}
By solving this equation, the mass of $\psi(4S)$ is estimated to be $4279.1$ MeV, which is consistent with calculations of $\psi(4S)$ within screened potential models \cite{Wang:2019mhs,Li:2009zu}, but larger than the PDG value of $4222.1 \pm 2.3$ MeV \cite{ParticleDataGroup:2024cfk}. From Table. \ref{psi4S and psi3D}, our results indicate that the total mass shift of the $\psi(4S)$ state is $\Delta M = -153.9$ MeV. The $\psi(4S)$ state strongly couples with the $DD^*$, $D^*D^*$, $DD_1(2430)^0$, $D^*D_1(2430)^0$, $D^*D_0(2300)$, and $D^*D_1(2420)$ channels.
\footnote{
Since the bare masses of $\psi(4S)$ and $\psi(3D)$ are approximately 4.5 GeV, the hadron loops involving $DD_1(2430)^0$, $DD_1(2420)$, $D^*D_1(2430)^0$, and $D^*D_1(2420)$ can contribute to the mass shifts of both $\psi(4S)$ and $\psi(3D)$. Here, the $D_1(2420)$ and $D_1(2430)^0$ share the same quantum numbers $J^{P}=1^+$. In the framework of heavy quark symmetry, the $J^{P}=1^+$ state can be expressed as a linear combination of the $1 ^1P_1$ and $1 ^3P_1$ states. The mixing scheme is described as follows:
\begin{gather}
	\begin{pmatrix}
		\mid D_1(2430)\rangle\\ \mid D_1(2420)\rangle
	\end{pmatrix}=
	\begin{pmatrix}
		\text{cos}\phi&\text{sin}\phi\\-\text{sin}\phi&\text{cos}\phi
	\end{pmatrix}
	\begin{pmatrix}
		\mid D(1 ^1P_1) \rangle\\ \mid D(1 ^3P_1) \rangle
	\end{pmatrix},
\end{gather}
where the mixing angle $\phi=-54.7^{\circ}$ can be determined in the heavy quark limit \cite{Cahn:2003cw}. For convenience, we use $DD$ to denote the sum of the $D\bar{D}$ and $\bar{D}D$ decay channels in the present work, and so forth. Here, the mass shift, mass and width are in units of MeV.}

Based on heavy quark spin symmetry (HQSS), the $D$ and $D^*$ mesons are degenerate in the heavy quark limit. This explains why the mass shift of $\psi(4S)$ receives similar contributions from the $DD_1(2430)^0$ and $D^*D_1(2430)^0$ channels. Similarly, the $D_1(2420)$ and $D^*_0(2300)$ mesons are also degenerate in the heavy quark limit. Therefore, the $D^*D_1(2420)$ and $D^*D^*_0(2300)$ channels also provide significant contributions to the mass shifts of the $\psi(4S)$ state. When $\psi(4S)$ couples to the $DD_1(2430)^0$, $D^*D_1(2430)^0$, and $D^*D_0(2300)$ channels through $S$-wave interactions, the line shape of $\Delta M$ exhibits three discontinuities. Additionally, the $DD^*$ and $D^*D^*$ channels also have significant branching ratios for mass shifts involving $\psi(4S)$, with their coupling primarily occurring via $P$-wave interactions. However, the coupling of $\psi(4S)$ with the $DD$ channel is relatively weak.

The $D_1$ state includes the $D_1(2420)$ with a narrow width of $31.3 \pm 1.9$ MeV and the $D_1(2430)^0$ with a broad width of $314 \pm 29$ MeV, as both $D_1(2420)$ and $D_1(2430)^0$ can decay into the $D^*\pi$ channel. Within the framework of HQSS, the $D_1(2420)$ decays to the $D^*\pi$ channel mainly via the $D$-wave, while the $D_1(2430)^0$ decays to the $D^*\pi$ channel primarily through the $S$-wave \cite{Li:2013yka}. Consequently, the coupling of $\psi(4S)$ with the $DD_1(2420)$ channel may be suppressed. On the other hand, the $D_1(2430)^0$ and $D_1(2420)$ are treated as mixed states in this study. Therefore, the coupling of $\psi(4S)$ with the $DD_1(2430)^0$ channel is stronger than that with the $DD_1(2420)$ channel. 

The predicted strong decay width of $\psi(4S)$ is $52.7$ MeV, with dominant decays into the $DD^*$ and $D^*D^*$ channels. The partial width ratio is predicted to be $\Gamma(DD^*):\Gamma(D^*D^*) =0.24 $. Contributions to the OZI-allowed two-body strong decay width from the $DD$, $D_sD_s$, $D_sD_s^*$, and $D_s^*D_s^*$ channels are relatively suppressed.
From the above analysis, the predicted mass of $\psi(4S)$ is approximately $60$ MeV larger than the measured mass of $\psi(4220)$, although the estimated two-body  OZI-allowed strong decay width of $\psi(4S)$ is close to the measured width of $\psi(4220)$. Therefore, further study is needed to better understand the nature of $\psi(4220)$.

By decomposing Eq. \eqref{mix-couple}, the coupled-channel equation for $\psi(3D)$ can be written as
\begin{eqnarray}\label{coupled-D}
M - M_D^0 - \Delta M_D(M) = 0.
\end{eqnarray}
The mass of $\psi(3D)$ is predicted to be $4338.5$ MeV based on the calculation in Eq. \eqref{coupled-D}. Our result is slightly smaller than the average experimental value of $4373 \pm 7$ MeV for $Y(4360)$ \cite{ParticleDataGroup:2024cfk}, but it is consistent with the predictions of screened potential models \cite{Wang:2019mhs,Li:2009zu}. From Fig. \ref{4S}, we find that the total mass shift of $\psi(3D)$ is estimated to be $-152.8$ MeV. The $D^*D^*$, $DD_1(2430)^0$, $DD_1(2420)$, $D^*D_1(2430)^0$ and $D^*D_1(2420)$ channels provide the dominant contributions to the mass shift of $\psi(3D)$. Among the coupled channels, $D^*D^*$ contributes the largest mass shift for $\psi(3D)$.

In the present work, the estimated width of $\psi(3D)$ is $23.6$ MeV, which is also smaller than the measured width of $118 \pm 12$ MeV for $Y(4360)$ \cite{ParticleDataGroup:2024cfk}.
 A similar conclusion is reached in the modified GI model of $\psi(3D)$ \cite{Wang:2019mhs}. Therefore, based on its decay properties, the $\psi(3D)$ charmonium state may be a narrow structure.

\subsection{The $4S$-$3D$ mixing scheme induced by coupled-channel effect}

According to our results, the calculated mass of the $\psi(4S)$ state is higher than the measured mass of the $\psi(4220)$. To resolve this mass discrepancy, we employ a comprehensive coupled-channel scheme, where the allowed hadronic loops composed of charmed mesons act as a bridge connecting the $\psi(4S)$ and $\psi(3D)$ states. This mechanism provides a source for the $4S$-$3D$ mixing.

\begin{figure}[!htbp]
	\centering
	\begin{minipage}[b]{0.48\textwidth}
		\centering
		\includegraphics[width=1\textwidth]{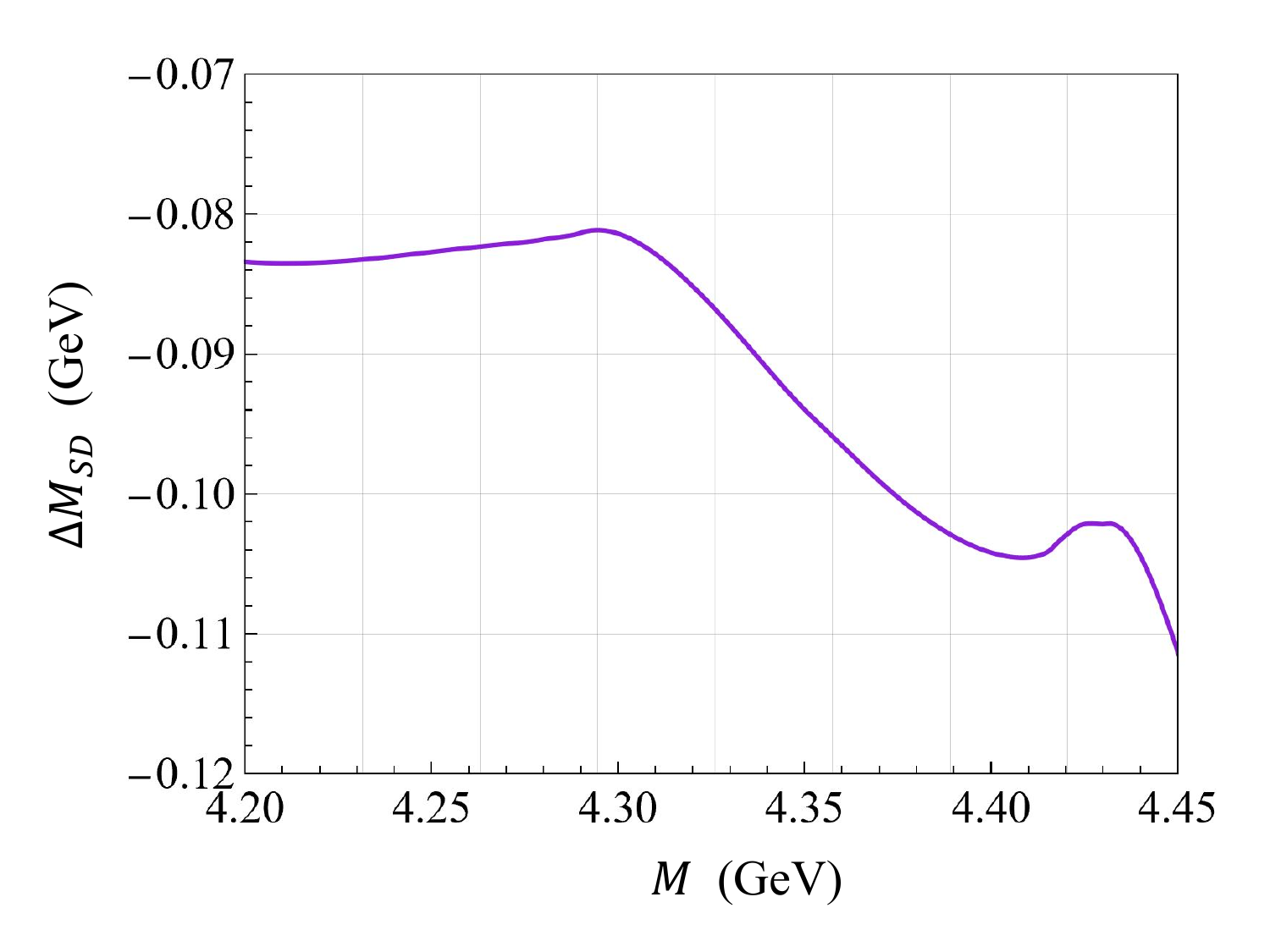}
	\end{minipage}
	\caption{The dependence of $\Delta M_{SD}$ on $M$.\label{4s3d}}
\end{figure}

In Fig. \ref{4s3d}, we illustrate the dependence of $\Delta M_{SD}$ on $M$. The result demonstrates that $\Delta M_{SD}$ is comparable in magnitude to the calculated values of $\Delta M_{S}$ and $\Delta M_{D}$. This observation strongly suggests the possibility of significant $4S$-$3D$ mixing. Our study explicitly confirms this behavior, highlighting the non-negligible contribution of such mixing effects.

The masses of the mixed states ${\psi}_{4S-3D}^\prime$ and ${\psi}_{4S-3D}^{\prime\prime}$, as well as the $4S$-$3D$ mixing angle, are determined by solving the reexpressed equation of Eq. \eqref{mix-couple}:
\begin{widetext}
\begin{eqnarray}
	\text{det}\begin{vmatrix}
		M_S^0 + \Delta M_{S}(M) - M &
        \langle \psi_S | H_{\text{T}} | \psi_D \rangle + \Delta M_{SD}(M) \\
		\langle \psi_D | H_{\text{T}} | \psi_S \rangle + \Delta M_{SD}(M)  &
        M_D^0 + \Delta M_{D}(M)- M
	\end{vmatrix} = 0.
 \label{haha}
\end{eqnarray}
\end{widetext}
In Table \ref{mix-decay}, we present the mass of the mixed state ${\psi}_{4S-3D}^\prime$, determined to be $4235.9$ MeV. This result is consistent with the PDG mass value of $4222.1 \pm 2.3$ MeV \cite{ParticleDataGroup:2024cfk} for the $Y(4220)$. Using Eq. \eqref{mix-couple}, the mixing angle between $\psi(4S)$ and $\psi(3D)$ is found to be 
\begin{equation*}
\theta = 35^\circ.
\end{equation*}
This value accounts for the large $4S$-$3D$ mixing in the $\psi(4220)$, as directly obtained from experimental data fitting in Ref. \cite{Wang:2019mhs}.

In addition to the mass, we investigate the decay properties of ${\psi}_{4S-3D}^\prime$. The estimated total decay width for the OZI-allowed two-body strong decay of ${\psi}_{4S-3D}^\prime$ is $32.0$ MeV, which aligns with the measured width of the $\psi(4220)$. We identify $D^*D^*$ as the dominant decay mode, with a branching fraction as high as $91.4\%$. These findings demonstrate that a comprehensive coupled-channel approach can successfully explain a large $4S$-$3D$ mixing angle and accurately reproduce both the mass and decay width of the $\psi(4220)$. Thus, interpreting the $\psi(4220)$ as a mixed state ${\psi}_{4S-3D}^{\prime}$ is well justified.

The estimated mass of another mixed state ${\psi}_{4S-3D}^{\prime\prime}$ is $4387.1$ MeV, consistent with the conclusion from a screening potential model \cite{Wang:2019mhs}. As shown in Table \ref{mix-decay}, the mixing angle $\theta = 35^\circ$ results in a predicted total width of two-body OZI-allowed strong decays to be $44.1$ MeV for ${\psi}_{4S-3D}^{\prime\prime}$, which indicates that it is possible to find this state in experiment. In fact, several former theoretical analyses of the processes $e^+e^-\to\psi(2S)\pi^+\pi^-$ \cite{Wang:2019mhs}, $e^+e^-\to K^+K^-J/\psi$ \cite{Wang:2022jxj}, $e^+e^-\to \pi^+D^0D^{*-}$ \cite{Wang:2023zxj}, and $e^+e^-\to \eta J/\psi$ \cite{Peng:2024xui} provide strong evidence for the existence of this predicted ${\psi}_{4S-3D}^{\prime\prime}$, which may be referred to as the $\psi(4380)$.

\begin{figure}[!htbp]
	\centering
	\begin{minipage}[b]{0.48\textwidth}
		\centering
		\includegraphics[width=1\textwidth]{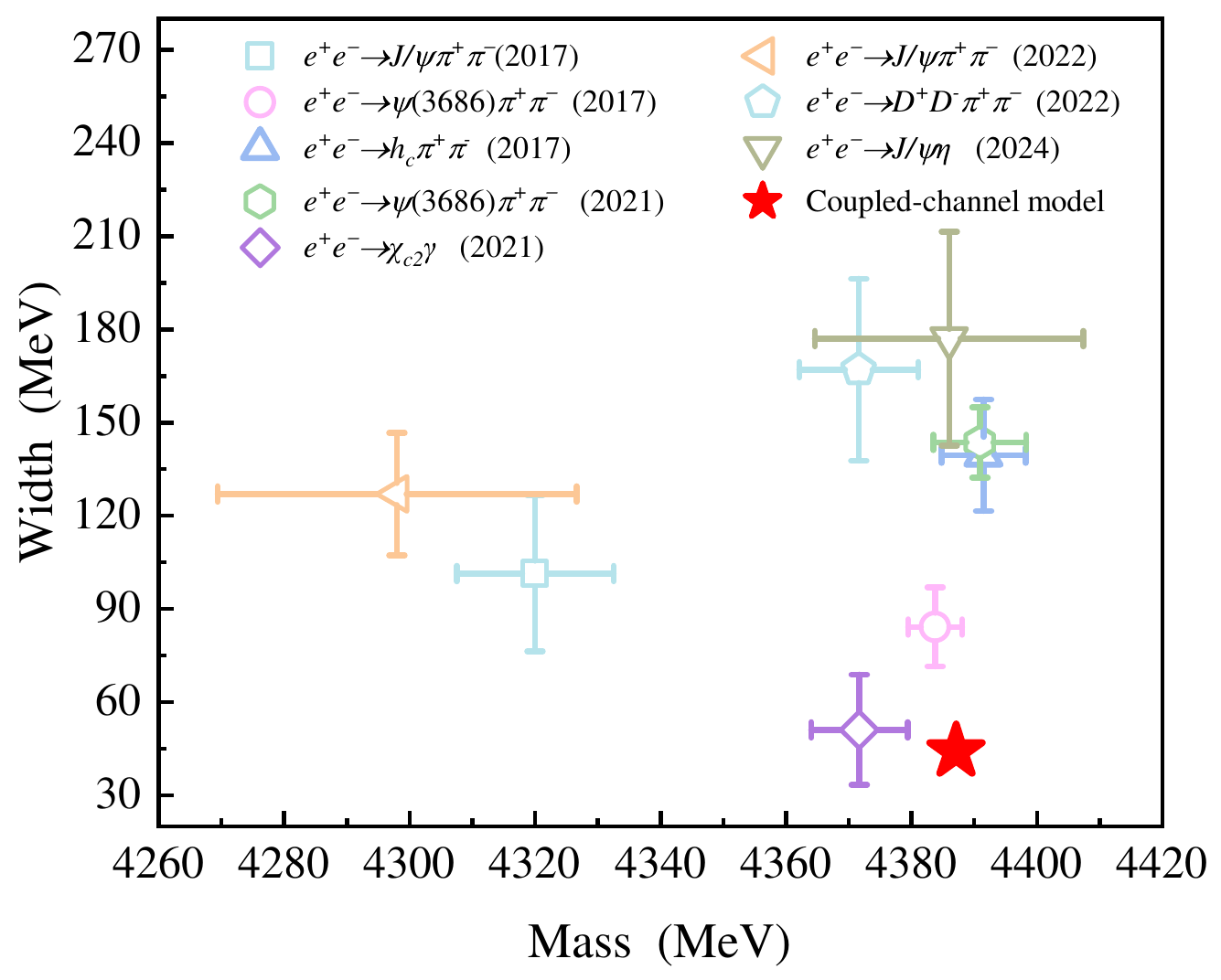}
	\end{minipage}
	\caption{The red pentagram denotes the calculated resonance parameters of ${\psi}_{4S-3D}^{\prime\prime}$ in our coupled-channel model. The other symbols represent the measured resonance parameters of vector charmonium-like structures around 4.36 GeV from various processes: $e^+e^-\to J/\psi\pi^+\pi^-$ \cite{BESIII:2016bnd}, $e^+e^-\to \psi(3686)\pi^+\pi^-$ \cite{BESIII:2017tqk}, $e^+e^-\to h_c\pi^+\pi^-$ \cite{BESIII:2016adj}, $e^+e^-\to \psi(3686)\pi^+\pi^-$ \cite{BESIII:2021njb}, $e^+e^-\to \chi_{c2}\gamma$ \cite{BESIII:2021yal}, $e^+e^-\to J/\psi\pi^+\pi^-$ \cite{BESIII:2022qal}, $e^+e^-\to D^+D^-\pi^+\pi^-$ \cite{BESIII:2022quc}, and $e^+e^-\to J/\psi\eta$ \cite{BESIII:2023tll}.\label{widthY4360}}
\end{figure}

In Fig. \ref{widthY4360}, we compare the resonance parameters of the mixed state ${\psi}_{4S-3D}^{\prime\prime}$ with those of the reported vector charmonium-like structures around $4.36$ GeV in different processes. Currently, the measured resonance parameters for vector charmonium-like structures in this energy region exhibit significant discrepancies. We find that the predicted resonance parameters of the mixed state ${\psi}_{4S-3D}^{\prime\prime}$ are consistent with the results from the $e^+e^-\to\chi_{c2}\gamma$ process, where the observed vector charmonium-like structure has a mass of $4371.7 \pm 7.5 \pm 1.8$ MeV and a width of $51.1 \pm 17.6 \pm 1.9$ MeV \cite{BESIII:2021yal}. However, more precise experimental data are required to conclusively determine its properties.

Our study indicates that the $4S$-$3D$ mixing scheme, induced by coupled-channel effects, can lead to significant differences in the dominant decay modes of ${\psi}_{4S-3D}^{\prime\prime}$ compared to those of a pure $\psi(3D)$ state. For the mixed state ${\psi}_{4S-3D}^{\prime\prime}$, the $DD^*_2(2460)$ channel plays a crucial role. Among the partial decay widths of the two-body OZI-allowed decays, the largest partial width originates from the $DD^*_2(2460)$ channel. Furthermore, our results show that the $D^*D^*$, $DD_1(2430)^0$, $DD_1(2420)$, and $DD^*_2(2460)$ channels are the dominant strong decay modes for ${\psi}_{4S-3D}^{\prime\prime}$, with the ratio $\Gamma(D^*D^*):\Gamma(DD_1(2430)^0):\Gamma(DD_1(2420)):\Gamma(DD^*_2(2460)) = 1.4:1:1.4:2.9$. These findings are also consistent with those reported in Ref. \cite{Wang:2019mhs}. 
Therefore, we encourage future experiments to search for ${\psi}_{4S-3D}^{\prime\prime}$ through the open-charm decay channels $D^*D^*$, $DD_1(2430)^0$, $DD_1(2420)$, and $DD^*_2(2460)$ and their subordinate decays.

As shown in Tables II and III, if the $\psi(4220)$ corresponds to the $\psi^{\prime}_{4S-3D}$ state in the $4S-3D$ mixing scheme, its decay width is narrower than that of $\psi(4S)$ because of smaller phase space.
The dominant decay channel of $\psi^{\prime}_{4S-3D}$ remains $D^*D^*$, with a branching ratio of 91.4\%, while the other decay modes such as $DD^{*}$ are significantly suppressed. 
If we treat $\psi(4380)$ as the higher mass state $\psi^{\prime\prime}_{4S-3D}$, its total decay width is significantly enhanced compared to that of $\psi(3D)$ state, primarily due to the larger phase space. Notably, the largest partial width of $\psi(4380)$ is $DD_2^*(2460)$ with a branching fraction reaching up to $37.3\%$, in sharp contrast to the prediction for the $\psi(3D)$, which has only about $0.9\%$.
This difference arises from the modified wave function induced by the $S-D$ mixing scheme. Additionally, $\psi(4380)$ also has dominant decay rates into the $D^*D^*$, $DD_1(2430)^0$, and $DD_1(2420)$ channels.
The predicted dominant decay channels can provide critical information to test the assignment of $\psi(4220)$ and support future experimental  searches for $\psi(4380)$.

\begin{table}[!htbp]
\renewcommand\arraystretch{1.5}
	\caption{Two-body strong decay widths of ${\psi}_{4S-3D}^\prime$ and ${\psi}_{4S-3D}^{\prime\prime}$ with mixing angle $35^{\circ}$. {The decay widths are in units of MeV. }}\centering
	\begin{tabular*}{1.0\columnwidth}{@{\extracolsep{\fill}}ccccc@{}}
    \toprule[1.00pt]
    \toprule[1.00pt]
    \label{mix-decay}
		&  \multicolumn{2}{c}{${\psi}_{4S-3D}^\prime$}  &\multicolumn{2}{c}{${\psi}_{4S-3D}^{\prime\prime}$}\\  
		&  \multicolumn{2}{c}{m=4235.9 MeV}  &\multicolumn{2}{c}{m=4387.1 MeV}\\ \hline 
		Channels &$\Gamma_{i}$&$\Gamma_{i}/\sum_i \Gamma_{i}$&$\Gamma_{{i}}$&$\Gamma_{i}/\sum_i \Gamma_{i}$\\ \hline 
$DD$              & 0.2      & 0.5\%  & 0.5        & 1.1\% \\
$DD^*$            & 1.7      & 5.4\%  & 0.7        & 1.7\% \\
$D_sD_s$          & 0.6      & 1.9\%  & 0.1        & 0.2\% \\
$D^*D^*$          & 29.2     & 91.4\% & 8.0        & 18.1\%\\
$D_sD_s^*$        & 0.0      & 0.0\%  & 2.5        & 5.7\% \\
$DD_0^*(2300)$    & $-$      & $-$   & $-$        & $-$  \\
$D_s^*D_s^*$      & 0.3      & 0.8\%  & 0.8        & 1.9\% \\
$DD_1(2430)^0$    & $-$      & $-$   & 5.7        & 13.0\%\\
$DD_1(2420)$      & $-$      & $-$   & 7.7        & 17.4\%\\
$D_sD_{s0}(2317)$ & $-$      & $-$   & $-$        & $-$  \\
$DD_2^*(2460)$    & $-$      & $-$   & 16.5       & 37.3\%\\
$D^*D_0^*(2300)$  & $-$      & $-$   & 1.6        & 3.7\% \\
Total      & 32.0     & 100\%  & 44.1       & 100\% \\\hline
   &$\Gamma_{\psi(4220)}=49\pm7$ \cite{ParticleDataGroup:2024cfk}  &   &  &  \\
       \bottomrule[1.00pt]
       \bottomrule[1.00pt]
	\end{tabular*}
\end{table}

\section{Summary}

In this work, we investigate the nature of the $\psi(4220)$ and its partner state $\psi(4380)$ within the framework of a coupled-channel model, focusing on the $4S$-$3D$ mixing scheme. By incorporating hadronic loop effects, we quantitatively analyze the mass shifts and decay properties of the $\psi(4S)$ and $\psi(3D)$ states. Our calculations reveal that the physical mass of the $\psi(4S)$ state is lowered to $4279.1$ MeV due to coupled-channel corrections, while the $\psi(3D)$ state is predicted to have a mass of $4338.5$ MeV. These results are consistent with predictions from screened potential models \cite{Li:2009zu,Wang:2019mhs} but highlight discrepancies with experimental measurements \cite{ParticleDataGroup:2024cfk}, particularly for the $\psi(4220)$.

To resolve the mass discrepancy, we propose a comprehensive $4S$-$3D$ mixing scheme induced by coupled-channel effects. The mixed state ${\psi}_{4S-3D}^\prime$, identified as the $\psi(4220)$, is found to have a mass of $4235.9$ MeV and a total decay width of $32.0$ MeV, in good agreement with experimental data \cite{ParticleDataGroup:2024cfk}. The mixing angle between $\psi(4S)$ and $\psi(3D)$ is determined to be $\theta = 35^\circ$, which confirms the large $4S$-$3D$ mixing suggested in Ref. \cite{Wang:2019mhs}. The dominant decay mode of ${\psi}_{4S-3D}^\prime$ is $D^*D^*$, with a branching fraction of $91.4\%$.

We also predict the existence of a second mixed state, ${\psi}_{4S-3D}^{\prime\prime}$, with a mass of $4387.1$ MeV and a total decay width of $44.1$ MeV. This state is suggested to correspond to the narrow $\psi(4380)$, which can be searched for in open-charm decay channels such as $D^*D^*$, $DD_1(2430)^0$, $DD_1(2420)$, and $DD^*_2(2460)$. Our results demonstrate that the large $4S$-$3D$ mixing angle is primarily driven by coupled-channel effects, providing a robust explanation for the observed properties of the $\psi(4220)$ and $\psi(4380)$.

This study underscores the importance of coupled-channel mechanisms in understanding the charmonium spectrum above 4 GeV and offers a new perspective on the role of meson loops in shaping the masses and decay properties of higher-lying charmonium states. Future experimental searches for the predicted $\psi(4380)$ state in the suggested decay channels will provide crucial tests of our theoretical framework.

{\bf Note added}: In preparing the present work, we observed that a mixing scheme induced by the coupled-channel effect was applied to study the bottomonium system in Ref. \cite{Ni:2025gvx} recently.

\section*{Acknowledgments}

This work is supported by the National Natural Science Foundation of China under Grant Nos. 12335001, 12247101, and 12405098,  the ‘111 Center’ under Grant No. B20063, the Natural Science Foundation of Gansu Province (No. 22JR5RA389, No. 25JRRA799), the fundamental Research Funds for the Central Universities, and the project for top-notch innovative talents of Gansu province.

\vfil

\bibliography{inport}

\begin{thebibliography}{94}%
\makeatletter
\providecommand \@ifxundefined [1]{%
 \@ifx{#1\undefined}
}%
\providecommand \@ifnum [1]{%
 \ifnum #1\expandafter \@firstoftwo
 \else \expandafter \@secondoftwo
 \fi
}%
\providecommand \@ifx [1]{%
 \ifx #1\expandafter \@firstoftwo
 \else \expandafter \@secondoftwo
 \fi
}%
\providecommand \natexlab [1]{#1}%
\providecommand \enquote  [1]{``#1''}%
\providecommand \bibnamefont  [1]{#1}%
\providecommand \bibfnamefont [1]{#1}%
\providecommand \citenamefont [1]{#1}%
\providecommand \href@noop [0]{\@secondoftwo}%
\providecommand \href [0]{\begingroup \@sanitize@url \@href}%
\providecommand \@href[1]{\@@startlink{#1}\@@href}%
\providecommand \@@href[1]{\endgroup#1\@@endlink}%
\providecommand \@sanitize@url [0]{\catcode `\\12\catcode `\$12\catcode
  `\&12\catcode `\#12\catcode `\^12\catcode `\_12\catcode `\%12\relax}%
\providecommand \@@startlink[1]{}%
\providecommand \@@endlink[0]{}%
\providecommand \url  [0]{\begingroup\@sanitize@url \@url }%
\providecommand \@url [1]{\endgroup\@href {#1}{\urlprefix }}%
\providecommand \urlprefix  [0]{URL }%
\providecommand \Eprint [0]{\href }%
\providecommand \doibase [0]{https://doi.org/}%
\providecommand \selectlanguage [0]{\@gobble}%
\providecommand \bibinfo  [0]{\@secondoftwo}%
\providecommand \bibfield  [0]{\@secondoftwo}%
\providecommand \translation [1]{[#1]}%
\providecommand \BibitemOpen [0]{}%
\providecommand \bibitemStop [0]{}%
\providecommand \bibitemNoStop [0]{.\EOS\space}%
\providecommand \EOS [0]{\spacefactor3000\relax}%
\providecommand \BibitemShut  [1]{\csname bibitem#1\endcsname}%
\let\auto@bib@innerbib\@empty
%</preamble>
\bibitem [{\citenamefont {Liu}(2014)}]{Liu:2013waa}%
  \BibitemOpen
  \bibfield  {author} {\bibinfo {author} {\bibfnamefont {X.}~\bibnamefont
  {Liu}},\ }\bibfield  {title} {\bibinfo {title} {{An overview of $XYZ$ new
  particles}},\ }\href {https://doi.org/10.1007/s11434-014-0407-2} {\bibfield
  {journal} {\bibinfo  {journal} {Chin. Sci. Bull.}\ }\textbf {\bibinfo
  {volume} {59}},\ \bibinfo {pages} {3815} (\bibinfo {year} {2014})},\ \Eprint
  {https://arxiv.org/abs/1312.7408} {arXiv:1312.7408 [hep-ph]} \BibitemShut
  {NoStop}%
\bibitem [{\citenamefont {Chen}\ \emph
  {et~al.}(2016{\natexlab{a}})\citenamefont {Chen}, \citenamefont {Chen},
  \citenamefont {Liu},\ and\ \citenamefont {Zhu}}]{Chen:2016qju}%
  \BibitemOpen
  \bibfield  {author} {\bibinfo {author} {\bibfnamefont {H.-X.}\ \bibnamefont
  {Chen}}, \bibinfo {author} {\bibfnamefont {W.}~\bibnamefont {Chen}}, \bibinfo
  {author} {\bibfnamefont {X.}~\bibnamefont {Liu}},\ and\ \bibinfo {author}
  {\bibfnamefont {S.-L.}\ \bibnamefont {Zhu}},\ }\bibfield  {title} {\bibinfo
  {title} {{The hidden-charm pentaquark and tetraquark states}},\ }\href
  {https://doi.org/10.1016/j.physrep.2016.05.004} {\bibfield  {journal}
  {\bibinfo  {journal} {Phys. Rept.}\ }\textbf {\bibinfo {volume} {639}},\
  \bibinfo {pages} {1} (\bibinfo {year} {2016}{\natexlab{a}})},\ \Eprint
  {https://arxiv.org/abs/1601.02092} {arXiv:1601.02092 [hep-ph]} \BibitemShut
  {NoStop}%
\bibitem [{\citenamefont {Chen}\ \emph {et~al.}(2017)\citenamefont {Chen},
  \citenamefont {Chen}, \citenamefont {Liu}, \citenamefont {Liu},\ and\
  \citenamefont {Zhu}}]{Chen:2016spr}%
  \BibitemOpen
  \bibfield  {author} {\bibinfo {author} {\bibfnamefont {H.-X.}\ \bibnamefont
  {Chen}}, \bibinfo {author} {\bibfnamefont {W.}~\bibnamefont {Chen}}, \bibinfo
  {author} {\bibfnamefont {X.}~\bibnamefont {Liu}}, \bibinfo {author}
  {\bibfnamefont {Y.-R.}\ \bibnamefont {Liu}},\ and\ \bibinfo {author}
  {\bibfnamefont {S.-L.}\ \bibnamefont {Zhu}},\ }\bibfield  {title} {\bibinfo
  {title} {{A review of the open charm and open bottom systems}},\ }\href
  {https://doi.org/10.1088/1361-6633/aa6420} {\bibfield  {journal} {\bibinfo
  {journal} {Rept. Prog. Phys.}\ }\textbf {\bibinfo {volume} {80}},\ \bibinfo
  {pages} {076201} (\bibinfo {year} {2017})},\ \Eprint
  {https://arxiv.org/abs/1609.08928} {arXiv:1609.08928 [hep-ph]} \BibitemShut
  {NoStop}%
\bibitem [{\citenamefont {Guo}\ \emph {et~al.}(2018)\citenamefont {Guo},
  \citenamefont {Hanhart}, \citenamefont {Mei\ss{}ner}, \citenamefont {Wang},
  \citenamefont {Zhao},\ and\ \citenamefont {Zou}}]{Guo:2017jvc}%
  \BibitemOpen
  \bibfield  {author} {\bibinfo {author} {\bibfnamefont {F.-K.}\ \bibnamefont
  {Guo}}, \bibinfo {author} {\bibfnamefont {C.}~\bibnamefont {Hanhart}},
  \bibinfo {author} {\bibfnamefont {U.-G.}\ \bibnamefont {Mei\ss{}ner}},
  \bibinfo {author} {\bibfnamefont {Q.}~\bibnamefont {Wang}}, \bibinfo {author}
  {\bibfnamefont {Q.}~\bibnamefont {Zhao}},\ and\ \bibinfo {author}
  {\bibfnamefont {B.-S.}\ \bibnamefont {Zou}},\ }\bibfield  {title} {\bibinfo
  {title} {{Hadronic molecules}},\ }\href
  {https://doi.org/10.1103/RevModPhys.90.015004} {\bibfield  {journal}
  {\bibinfo  {journal} {Rev. Mod. Phys.}\ }\textbf {\bibinfo {volume} {90}},\
  \bibinfo {pages} {015004} (\bibinfo {year} {2018})},\ \bibinfo {note}
  {[Erratum: Rev.Mod.Phys. 94, 029901 (2022)]},\ \Eprint
  {https://arxiv.org/abs/1705.00141} {arXiv:1705.00141 [hep-ph]} \BibitemShut
  {NoStop}%
\bibitem [{\citenamefont {Liu}\ \emph {et~al.}(2019)\citenamefont {Liu},
  \citenamefont {Chen}, \citenamefont {Chen}, \citenamefont {Liu},\ and\
  \citenamefont {Zhu}}]{Liu:2019zoy}%
  \BibitemOpen
  \bibfield  {author} {\bibinfo {author} {\bibfnamefont {Y.-R.}\ \bibnamefont
  {Liu}}, \bibinfo {author} {\bibfnamefont {H.-X.}\ \bibnamefont {Chen}},
  \bibinfo {author} {\bibfnamefont {W.}~\bibnamefont {Chen}}, \bibinfo {author}
  {\bibfnamefont {X.}~\bibnamefont {Liu}},\ and\ \bibinfo {author}
  {\bibfnamefont {S.-L.}\ \bibnamefont {Zhu}},\ }\bibfield  {title} {\bibinfo
  {title} {{Pentaquark and Tetraquark states}},\ }\href
  {https://doi.org/10.1016/j.ppnp.2019.04.003} {\bibfield  {journal} {\bibinfo
  {journal} {Prog. Part. Nucl. Phys.}\ }\textbf {\bibinfo {volume} {107}},\
  \bibinfo {pages} {237} (\bibinfo {year} {2019})},\ \Eprint
  {https://arxiv.org/abs/1903.11976} {arXiv:1903.11976 [hep-ph]} \BibitemShut
  {NoStop}%
\bibitem [{\citenamefont {Brambilla}\ \emph {et~al.}(2020)\citenamefont
  {Brambilla}, \citenamefont {Eidelman}, \citenamefont {Hanhart}, \citenamefont
  {Nefediev}, \citenamefont {Shen}, \citenamefont {Thomas}, \citenamefont
  {Vairo},\ and\ \citenamefont {Yuan}}]{Brambilla:2019esw}%
  \BibitemOpen
  \bibfield  {author} {\bibinfo {author} {\bibfnamefont {N.}~\bibnamefont
  {Brambilla}}, \bibinfo {author} {\bibfnamefont {S.}~\bibnamefont {Eidelman}},
  \bibinfo {author} {\bibfnamefont {C.}~\bibnamefont {Hanhart}}, \bibinfo
  {author} {\bibfnamefont {A.}~\bibnamefont {Nefediev}}, \bibinfo {author}
  {\bibfnamefont {C.-P.}\ \bibnamefont {Shen}}, \bibinfo {author}
  {\bibfnamefont {C.~E.}\ \bibnamefont {Thomas}}, \bibinfo {author}
  {\bibfnamefont {A.}~\bibnamefont {Vairo}},\ and\ \bibinfo {author}
  {\bibfnamefont {C.-Z.}\ \bibnamefont {Yuan}},\ }\bibfield  {title} {\bibinfo
  {title} {{The $XYZ$ states: experimental and theoretical status and
  perspectives}},\ }\href {https://doi.org/10.1016/j.physrep.2020.05.001}
  {\bibfield  {journal} {\bibinfo  {journal} {Phys. Rept.}\ }\textbf {\bibinfo
  {volume} {873}},\ \bibinfo {pages} {1} (\bibinfo {year} {2020})},\ \Eprint
  {https://arxiv.org/abs/1907.07583} {arXiv:1907.07583 [hep-ex]} \BibitemShut
  {NoStop}%
\bibitem [{\citenamefont {Chen}\ \emph {et~al.}(2023)\citenamefont {Chen},
  \citenamefont {Chen}, \citenamefont {Liu}, \citenamefont {Liu},\ and\
  \citenamefont {Zhu}}]{Chen:2022asf}%
  \BibitemOpen
  \bibfield  {author} {\bibinfo {author} {\bibfnamefont {H.-X.}\ \bibnamefont
  {Chen}}, \bibinfo {author} {\bibfnamefont {W.}~\bibnamefont {Chen}}, \bibinfo
  {author} {\bibfnamefont {X.}~\bibnamefont {Liu}}, \bibinfo {author}
  {\bibfnamefont {Y.-R.}\ \bibnamefont {Liu}},\ and\ \bibinfo {author}
  {\bibfnamefont {S.-L.}\ \bibnamefont {Zhu}},\ }\bibfield  {title} {\bibinfo
  {title} {{An updated review of the new hadron states}},\ }\href
  {https://doi.org/10.1088/1361-6633/aca3b6} {\bibfield  {journal} {\bibinfo
  {journal} {Rept. Prog. Phys.}\ }\textbf {\bibinfo {volume} {86}},\ \bibinfo
  {pages} {026201} (\bibinfo {year} {2023})},\ \Eprint
  {https://arxiv.org/abs/2204.02649} {arXiv:2204.02649 [hep-ph]} \BibitemShut
  {NoStop}%
\bibitem [{\citenamefont {Ablikim}\ \emph {et~al.}(2020)\citenamefont {Ablikim}
  \emph {et~al.}}]{BESIII:2020nme}%
  \BibitemOpen
  \bibfield  {author} {\bibinfo {author} {\bibfnamefont {M.}~\bibnamefont
  {Ablikim}} \emph {et~al.} (\bibinfo {collaboration} {BESIII}),\ }\bibfield
  {title} {\bibinfo {title} {{Future Physics Programme of BESIII}},\ }\href
  {https://doi.org/10.1088/1674-1137/44/4/040001} {\bibfield  {journal}
  {\bibinfo  {journal} {Chin. Phys. C}\ }\textbf {\bibinfo {volume} {44}},\
  \bibinfo {pages} {040001} (\bibinfo {year} {2020})},\ \Eprint
  {https://arxiv.org/abs/1912.05983} {arXiv:1912.05983 [hep-ex]} \BibitemShut
  {NoStop}%
\bibitem [{\citenamefont {Aubert}\ \emph {et~al.}(2005)\citenamefont {Aubert}
  \emph {et~al.}}]{BaBar:2005hhc}%
  \BibitemOpen
  \bibfield  {author} {\bibinfo {author} {\bibfnamefont {B.}~\bibnamefont
  {Aubert}} \emph {et~al.} (\bibinfo {collaboration} {BaBar}),\ }\bibfield
  {title} {\bibinfo {title} {{Observation of a broad structure in the $\pi^+
  \pi^- J/\psi$ mass spectrum around 4.26-GeV/c$^2$}},\ }\href
  {https://doi.org/10.1103/PhysRevLett.95.142001} {\bibfield  {journal}
  {\bibinfo  {journal} {Phys. Rev. Lett.}\ }\textbf {\bibinfo {volume} {95}},\
  \bibinfo {pages} {142001} (\bibinfo {year} {2005})},\ \Eprint
  {https://arxiv.org/abs/hep-ex/0506081} {arXiv:hep-ex/0506081} \BibitemShut
  {NoStop}%
\bibitem [{\citenamefont {He}\ \emph {et~al.}(2006)\citenamefont {He} \emph
  {et~al.}}]{CLEO:2006tct}%
  \BibitemOpen
  \bibfield  {author} {\bibinfo {author} {\bibfnamefont {Q.}~\bibnamefont {He}}
  \emph {et~al.} (\bibinfo {collaboration} {CLEO}),\ }\bibfield  {title}
  {\bibinfo {title} {{Confirmation of the Y(4260) resonance production in
  ISR}},\ }\href {https://doi.org/10.1103/PhysRevD.74.091104} {\bibfield
  {journal} {\bibinfo  {journal} {Phys. Rev. D}\ }\textbf {\bibinfo {volume}
  {74}},\ \bibinfo {pages} {091104} (\bibinfo {year} {2006})},\ \Eprint
  {https://arxiv.org/abs/hep-ex/0611021} {arXiv:hep-ex/0611021} \BibitemShut
  {NoStop}%
\bibitem [{\citenamefont {Yuan}\ \emph {et~al.}(2007)\citenamefont {Yuan} \emph
  {et~al.}}]{Belle:2007dxy}%
  \BibitemOpen
  \bibfield  {author} {\bibinfo {author} {\bibfnamefont {C.~Z.}\ \bibnamefont
  {Yuan}} \emph {et~al.} (\bibinfo {collaboration} {Belle}),\ }\bibfield
  {title} {\bibinfo {title} {{Measurement of $e^+ e^-\to \pi^+ \pi^- J/\psi$
  cross-section via initial state radiation at Belle}},\ }\href
  {https://doi.org/10.1103/PhysRevLett.99.182004} {\bibfield  {journal}
  {\bibinfo  {journal} {Phys. Rev. Lett.}\ }\textbf {\bibinfo {volume} {99}},\
  \bibinfo {pages} {182004} (\bibinfo {year} {2007})},\ \Eprint
  {https://arxiv.org/abs/0707.2541} {arXiv:0707.2541 [hep-ex]} \BibitemShut
  {NoStop}%
\bibitem [{\citenamefont {Zhu}(2005)}]{Zhu:2005hp}%
  \BibitemOpen
  \bibfield  {author} {\bibinfo {author} {\bibfnamefont {S.-L.}\ \bibnamefont
  {Zhu}},\ }\bibfield  {title} {\bibinfo {title} {{The Possible interpretations
  of Y(4260)}},\ }\href {https://doi.org/10.1016/j.physletb.2005.08.068}
  {\bibfield  {journal} {\bibinfo  {journal} {Phys. Lett. B}\ }\textbf
  {\bibinfo {volume} {625}},\ \bibinfo {pages} {212} (\bibinfo {year}
  {2005})},\ \Eprint {https://arxiv.org/abs/hep-ph/0507025}
  {arXiv:hep-ph/0507025} \BibitemShut {NoStop}%
\bibitem [{\citenamefont {Kou}\ and\ \citenamefont {Pene}(2005)}]{Kou:2005gt}%
  \BibitemOpen
  \bibfield  {author} {\bibinfo {author} {\bibfnamefont {E.}~\bibnamefont
  {Kou}}\ and\ \bibinfo {author} {\bibfnamefont {O.}~\bibnamefont {Pene}},\
  }\bibfield  {title} {\bibinfo {title} {{Suppressed decay into open charm for
  the Y(4260) being an hybrid}},\ }\href
  {https://doi.org/10.1016/j.physletb.2005.09.013} {\bibfield  {journal}
  {\bibinfo  {journal} {Phys. Lett. B}\ }\textbf {\bibinfo {volume} {631}},\
  \bibinfo {pages} {164} (\bibinfo {year} {2005})},\ \Eprint
  {https://arxiv.org/abs/hep-ph/0507119} {arXiv:hep-ph/0507119} \BibitemShut
  {NoStop}%
\bibitem [{\citenamefont {Maiani}\ \emph {et~al.}(2005)\citenamefont {Maiani},
  \citenamefont {Riquer}, \citenamefont {Piccinini},\ and\ \citenamefont
  {Polosa}}]{Maiani:2005pe}%
  \BibitemOpen
  \bibfield  {author} {\bibinfo {author} {\bibfnamefont {L.}~\bibnamefont
  {Maiani}}, \bibinfo {author} {\bibfnamefont {V.}~\bibnamefont {Riquer}},
  \bibinfo {author} {\bibfnamefont {F.}~\bibnamefont {Piccinini}},\ and\
  \bibinfo {author} {\bibfnamefont {A.~D.}\ \bibnamefont {Polosa}},\ }\bibfield
   {title} {\bibinfo {title} {{Four quark interpretation of Y(4260)}},\ }\href
  {https://doi.org/10.1103/PhysRevD.72.031502} {\bibfield  {journal} {\bibinfo
  {journal} {Phys. Rev. D}\ }\textbf {\bibinfo {volume} {72}},\ \bibinfo
  {pages} {031502} (\bibinfo {year} {2005})},\ \Eprint
  {https://arxiv.org/abs/hep-ph/0507062} {arXiv:hep-ph/0507062} \BibitemShut
  {NoStop}%
\bibitem [{\citenamefont {Ebert}\ \emph {et~al.}(2008)\citenamefont {Ebert},
  \citenamefont {Faustov},\ and\ \citenamefont {Galkin}}]{Ebert:2008kb}%
  \BibitemOpen
  \bibfield  {author} {\bibinfo {author} {\bibfnamefont {D.}~\bibnamefont
  {Ebert}}, \bibinfo {author} {\bibfnamefont {R.~N.}\ \bibnamefont {Faustov}},\
  and\ \bibinfo {author} {\bibfnamefont {V.~O.}\ \bibnamefont {Galkin}},\
  }\bibfield  {title} {\bibinfo {title} {{Excited heavy tetraquarks with hidden
  charm}},\ }\href {https://doi.org/10.1140/epjc/s10052-008-0754-8} {\bibfield
  {journal} {\bibinfo  {journal} {Eur. Phys. J. C}\ }\textbf {\bibinfo {volume}
  {58}},\ \bibinfo {pages} {399} (\bibinfo {year} {2008})},\ \Eprint
  {https://arxiv.org/abs/0808.3912} {arXiv:0808.3912 [hep-ph]} \BibitemShut
  {NoStop}%
\bibitem [{\citenamefont {Chen}\ \emph
  {et~al.}(2015{\natexlab{a}})\citenamefont {Chen}, \citenamefont {Maiani},
  \citenamefont {Polosa},\ and\ \citenamefont {Riquer}}]{Chen:2015dig}%
  \BibitemOpen
  \bibfield  {author} {\bibinfo {author} {\bibfnamefont {H.~X.}\ \bibnamefont
  {Chen}}, \bibinfo {author} {\bibfnamefont {L.}~\bibnamefont {Maiani}},
  \bibinfo {author} {\bibfnamefont {A.~D.}\ \bibnamefont {Polosa}},\ and\
  \bibinfo {author} {\bibfnamefont {V.}~\bibnamefont {Riquer}},\ }\bibfield
  {title} {\bibinfo {title} {{$Y(4260)\rightarrow \gamma + X(3872)$ in the
  diquarkonium picture}},\ }\href
  {https://doi.org/10.1140/epjc/s10052-015-3781-2} {\bibfield  {journal}
  {\bibinfo  {journal} {Eur. Phys. J. C}\ }\textbf {\bibinfo {volume} {75}},\
  \bibinfo {pages} {550} (\bibinfo {year} {2015}{\natexlab{a}})},\ \Eprint
  {https://arxiv.org/abs/1510.03626} {arXiv:1510.03626 [hep-ph]} \BibitemShut
  {NoStop}%
\bibitem [{\citenamefont {Chen}\ and\ \citenamefont {Zhu}(2011)}]{Chen:2010ze}%
  \BibitemOpen
  \bibfield  {author} {\bibinfo {author} {\bibfnamefont {W.}~\bibnamefont
  {Chen}}\ and\ \bibinfo {author} {\bibfnamefont {S.-L.}\ \bibnamefont {Zhu}},\
  }\bibfield  {title} {\bibinfo {title} {{The Vector and Axial-Vector
  Charmonium-like States}},\ }\href
  {https://doi.org/10.1103/PhysRevD.83.034010} {\bibfield  {journal} {\bibinfo
  {journal} {Phys. Rev. D}\ }\textbf {\bibinfo {volume} {83}},\ \bibinfo
  {pages} {034010} (\bibinfo {year} {2011})},\ \Eprint
  {https://arxiv.org/abs/1010.3397} {arXiv:1010.3397 [hep-ph]} \BibitemShut
  {NoStop}%
\bibitem [{\citenamefont {Zhang}\ and\ \citenamefont
  {Huang}(2011)}]{Zhang:2010mw}%
  \BibitemOpen
  \bibfield  {author} {\bibinfo {author} {\bibfnamefont {J.-R.}\ \bibnamefont
  {Zhang}}\ and\ \bibinfo {author} {\bibfnamefont {M.-Q.}\ \bibnamefont
  {Huang}},\ }\bibfield  {title} {\bibinfo {title} {{The $P$-wave
  $[cs][\bar{c}\bar{s}]$ tetraquark state: $Y(4260)$ or $Y(4660)$?}},\ }\href
  {https://doi.org/10.1103/PhysRevD.83.036005} {\bibfield  {journal} {\bibinfo
  {journal} {Phys. Rev. D}\ }\textbf {\bibinfo {volume} {83}},\ \bibinfo
  {pages} {036005} (\bibinfo {year} {2011})},\ \Eprint
  {https://arxiv.org/abs/1011.2818} {arXiv:1011.2818 [hep-ph]} \BibitemShut
  {NoStop}%
\bibitem [{\citenamefont {Yuan}\ \emph {et~al.}(2006)\citenamefont {Yuan},
  \citenamefont {Wang},\ and\ \citenamefont {Mo}}]{Yuan:2005dr}%
  \BibitemOpen
  \bibfield  {author} {\bibinfo {author} {\bibfnamefont {C.~Z.}\ \bibnamefont
  {Yuan}}, \bibinfo {author} {\bibfnamefont {P.}~\bibnamefont {Wang}},\ and\
  \bibinfo {author} {\bibfnamefont {X.~H.}\ \bibnamefont {Mo}},\ }\bibfield
  {title} {\bibinfo {title} {{The Y(4260) as an omega chi(c1) molecular
  state}},\ }\href {https://doi.org/10.1016/j.physletb.2006.01.031} {\bibfield
  {journal} {\bibinfo  {journal} {Phys. Lett. B}\ }\textbf {\bibinfo {volume}
  {634}},\ \bibinfo {pages} {399} (\bibinfo {year} {2006})},\ \Eprint
  {https://arxiv.org/abs/hep-ph/0511107} {arXiv:hep-ph/0511107} \BibitemShut
  {NoStop}%
\bibitem [{\citenamefont {Cleven}\ \emph {et~al.}(2014)\citenamefont {Cleven},
  \citenamefont {Wang}, \citenamefont {Guo}, \citenamefont {Hanhart},
  \citenamefont {Mei\ss{}ner},\ and\ \citenamefont {Zhao}}]{Cleven:2013mka}%
  \BibitemOpen
  \bibfield  {author} {\bibinfo {author} {\bibfnamefont {M.}~\bibnamefont
  {Cleven}}, \bibinfo {author} {\bibfnamefont {Q.}~\bibnamefont {Wang}},
  \bibinfo {author} {\bibfnamefont {F.-K.}\ \bibnamefont {Guo}}, \bibinfo
  {author} {\bibfnamefont {C.}~\bibnamefont {Hanhart}}, \bibinfo {author}
  {\bibfnamefont {U.-G.}\ \bibnamefont {Mei\ss{}ner}},\ and\ \bibinfo {author}
  {\bibfnamefont {Q.}~\bibnamefont {Zhao}},\ }\bibfield  {title} {\bibinfo
  {title} {{$Y(4260)$ as the first $S$-wave open charm vector molecular
  state?}},\ }\href {https://doi.org/10.1103/PhysRevD.90.074039} {\bibfield
  {journal} {\bibinfo  {journal} {Phys. Rev. D}\ }\textbf {\bibinfo {volume}
  {90}},\ \bibinfo {pages} {074039} (\bibinfo {year} {2014})},\ \Eprint
  {https://arxiv.org/abs/1310.2190} {arXiv:1310.2190 [hep-ph]} \BibitemShut
  {NoStop}%
\bibitem [{\citenamefont {Close}\ \emph {et~al.}(2010)\citenamefont {Close},
  \citenamefont {Downum},\ and\ \citenamefont {Thomas}}]{Close:2010wq}%
  \BibitemOpen
  \bibfield  {author} {\bibinfo {author} {\bibfnamefont {F.}~\bibnamefont
  {Close}}, \bibinfo {author} {\bibfnamefont {C.}~\bibnamefont {Downum}},\ and\
  \bibinfo {author} {\bibfnamefont {C.~E.}\ \bibnamefont {Thomas}},\ }\bibfield
   {title} {\bibinfo {title} {{Novel Charmonium and Bottomonium Spectroscopies
  due to Deeply Bound Hadronic Molecules from Single Pion Exchange}},\ }\href
  {https://doi.org/10.1103/PhysRevD.81.074033} {\bibfield  {journal} {\bibinfo
  {journal} {Phys. Rev. D}\ }\textbf {\bibinfo {volume} {81}},\ \bibinfo
  {pages} {074033} (\bibinfo {year} {2010})},\ \Eprint
  {https://arxiv.org/abs/1001.2553} {arXiv:1001.2553 [hep-ph]} \BibitemShut
  {NoStop}%
\bibitem [{\citenamefont {Li}\ and\ \citenamefont {Liu}(2013)}]{Li:2013yla}%
  \BibitemOpen
  \bibfield  {author} {\bibinfo {author} {\bibfnamefont {G.}~\bibnamefont
  {Li}}\ and\ \bibinfo {author} {\bibfnamefont {X.-H.}\ \bibnamefont {Liu}},\
  }\bibfield  {title} {\bibinfo {title} {{Investigating possible decay modes of
  Y(4260) under $D_1(2420)\bar{D}$ + c.c. molecular state ansatz}},\ }\href
  {https://doi.org/10.1103/PhysRevD.88.094008} {\bibfield  {journal} {\bibinfo
  {journal} {Phys. Rev. D}\ }\textbf {\bibinfo {volume} {88}},\ \bibinfo
  {pages} {094008} (\bibinfo {year} {2013})},\ \Eprint
  {https://arxiv.org/abs/1307.2622} {arXiv:1307.2622 [hep-ph]} \BibitemShut
  {NoStop}%
\bibitem [{\citenamefont {Lu}\ \emph {et~al.}(2017)\citenamefont {Lu},
  \citenamefont {Anwar},\ and\ \citenamefont {Zou}}]{Lu:2017yhl}%
  \BibitemOpen
  \bibfield  {author} {\bibinfo {author} {\bibfnamefont {Y.}~\bibnamefont
  {Lu}}, \bibinfo {author} {\bibfnamefont {M.~N.}\ \bibnamefont {Anwar}},\ and\
  \bibinfo {author} {\bibfnamefont {B.-S.}\ \bibnamefont {Zou}},\ }\bibfield
  {title} {\bibinfo {title} {{$X(4260)$ Revisited: A Coupled Channel
  Perspective}},\ }\href {https://doi.org/10.1103/PhysRevD.96.114022}
  {\bibfield  {journal} {\bibinfo  {journal} {Phys. Rev. D}\ }\textbf {\bibinfo
  {volume} {96}},\ \bibinfo {pages} {114022} (\bibinfo {year} {2017})},\
  \Eprint {https://arxiv.org/abs/1705.00449} {arXiv:1705.00449 [hep-ph]}
  \BibitemShut {NoStop}%
\bibitem [{\citenamefont {Abe}\ \emph {et~al.}(2007)\citenamefont {Abe} \emph
  {et~al.}}]{Belle:2006hvs}%
  \BibitemOpen
  \bibfield  {author} {\bibinfo {author} {\bibfnamefont {K.}~\bibnamefont
  {Abe}} \emph {et~al.} (\bibinfo {collaboration} {Belle}),\ }\bibfield
  {title} {\bibinfo {title} {{Measurement of the near-threshold$e^+ e^-\to
  D^{(*)\pm} D^{(*)\mp} $cross section using initial-state radiation}},\ }\href
  {https://doi.org/10.1103/PhysRevLett.98.092001} {\bibfield  {journal}
  {\bibinfo  {journal} {Phys. Rev. Lett.}\ }\textbf {\bibinfo {volume} {98}},\
  \bibinfo {pages} {092001} (\bibinfo {year} {2007})},\ \Eprint
  {https://arxiv.org/abs/hep-ex/0608018} {arXiv:hep-ex/0608018} \BibitemShut
  {NoStop}%
\bibitem [{\citenamefont {Pakhlova}\ \emph
  {et~al.}(2008{\natexlab{a}})\citenamefont {Pakhlova} \emph
  {et~al.}}]{Belle:2007qxm}%
  \BibitemOpen
  \bibfield  {author} {\bibinfo {author} {\bibfnamefont {G.}~\bibnamefont
  {Pakhlova}} \emph {et~al.} (\bibinfo {collaboration} {Belle}),\ }\bibfield
  {title} {\bibinfo {title} {{Measurement of the near-threshold $e^+ e^-\to D
  \bar{D}$ cross section using initial-state radiation}},\ }\href
  {https://doi.org/10.1103/PhysRevD.77.011103} {\bibfield  {journal} {\bibinfo
  {journal} {Phys. Rev. D}\ }\textbf {\bibinfo {volume} {77}},\ \bibinfo
  {pages} {011103} (\bibinfo {year} {2008}{\natexlab{a}})},\ \Eprint
  {https://arxiv.org/abs/0708.0082} {arXiv:0708.0082 [hep-ex]} \BibitemShut
  {NoStop}%
\bibitem [{\citenamefont {Pakhlova}\ \emph
  {et~al.}(2008{\natexlab{b}})\citenamefont {Pakhlova} \emph
  {et~al.}}]{Belle:2007xvy}%
  \BibitemOpen
  \bibfield  {author} {\bibinfo {author} {\bibfnamefont {G.}~\bibnamefont
  {Pakhlova}} \emph {et~al.} (\bibinfo {collaboration} {Belle}),\ }\bibfield
  {title} {\bibinfo {title} {{Observation of $\psi(4415)\to D
  \bar{D^*_2}(2460)$ decay using initial-state radiation}},\ }\href
  {https://doi.org/10.1103/PhysRevLett.100.062001} {\bibfield  {journal}
  {\bibinfo  {journal} {Phys. Rev. Lett.}\ }\textbf {\bibinfo {volume} {100}},\
  \bibinfo {pages} {062001} (\bibinfo {year} {2008}{\natexlab{b}})},\ \Eprint
  {https://arxiv.org/abs/0708.3313} {arXiv:0708.3313 [hep-ex]} \BibitemShut
  {NoStop}%
\bibitem [{\citenamefont {Pakhlova}\ \emph {et~al.}(2009)\citenamefont
  {Pakhlova} \emph {et~al.}}]{Belle:2009dus}%
  \BibitemOpen
  \bibfield  {author} {\bibinfo {author} {\bibfnamefont {G.}~\bibnamefont
  {Pakhlova}} \emph {et~al.} (\bibinfo {collaboration} {Belle}),\ }\bibfield
  {title} {\bibinfo {title} {{Measurement of the $e^+ e^-\to D^0 D^{*-} \pi^+$
  cross section using initial-state radiation}},\ }\href
  {https://doi.org/10.1103/PhysRevD.80.091101} {\bibfield  {journal} {\bibinfo
  {journal} {Phys. Rev. D}\ }\textbf {\bibinfo {volume} {80}},\ \bibinfo
  {pages} {091101} (\bibinfo {year} {2009})},\ \Eprint
  {https://arxiv.org/abs/0908.0231} {arXiv:0908.0231 [hep-ex]} \BibitemShut
  {NoStop}%
\bibitem [{\citenamefont {Cronin-Hennessy}\ \emph {et~al.}(2009)\citenamefont
  {Cronin-Hennessy} \emph {et~al.}}]{CLEO:2008ojp}%
  \BibitemOpen
  \bibfield  {author} {\bibinfo {author} {\bibfnamefont {D.}~\bibnamefont
  {Cronin-Hennessy}} \emph {et~al.} (\bibinfo {collaboration} {CLEO}),\
  }\bibfield  {title} {\bibinfo {title} {{Measurement of Charm Production Cross
  Sections in $e^+e^-$Annihilation at Energies between 3.97 and 4.26 GeV}},\
  }\href {https://doi.org/10.1103/PhysRevD.80.072001} {\bibfield  {journal}
  {\bibinfo  {journal} {Phys. Rev. D}\ }\textbf {\bibinfo {volume} {80}},\
  \bibinfo {pages} {072001} (\bibinfo {year} {2009})},\ \Eprint
  {https://arxiv.org/abs/0801.3418} {arXiv:0801.3418 [hep-ex]} \BibitemShut
  {NoStop}%
\bibitem [{\citenamefont {Ablikim}\ \emph {et~al.}(2009)\citenamefont {Ablikim}
  \emph {et~al.}}]{BES:2009ejh}%
  \BibitemOpen
  \bibfield  {author} {\bibinfo {author} {\bibfnamefont {M.}~\bibnamefont
  {Ablikim}} \emph {et~al.} (\bibinfo {collaboration} {BES}),\ }\bibfield
  {title} {\bibinfo {title} {{R value measurements for $e^+ e^-$ annihilation
  at 2.60 GeV, 3.07 GeV and 3.65 GeV}},\ }\href
  {https://doi.org/10.1016/j.physletb.2009.05.055} {\bibfield  {journal}
  {\bibinfo  {journal} {Phys. Lett. B}\ }\textbf {\bibinfo {volume} {677}},\
  \bibinfo {pages} {239} (\bibinfo {year} {2009})},\ \Eprint
  {https://arxiv.org/abs/0903.0900} {arXiv:0903.0900 [hep-ex]} \BibitemShut
  {NoStop}%
\bibitem [{\citenamefont {Burmester}\ \emph {et~al.}(1977)\citenamefont
  {Burmester} \emph {et~al.}}]{PLUTO:1976jbe}%
  \BibitemOpen
  \bibfield  {author} {\bibinfo {author} {\bibfnamefont {J.}~\bibnamefont
  {Burmester}} \emph {et~al.} (\bibinfo {collaboration} {PLUTO}),\ }\bibfield
  {title} {\bibinfo {title} {{The Total Hadronic Cross-Section for $e^+ e^-$
  Annihilation Between 3.1 GeV and 4.8 GeV Center-Of-Mass Energy}},\ }\href
  {https://doi.org/10.1016/0370-2693(77)90023-5} {\bibfield  {journal}
  {\bibinfo  {journal} {Phys. Lett. B}\ }\textbf {\bibinfo {volume} {66}},\
  \bibinfo {pages} {395} (\bibinfo {year} {1977})}\BibitemShut {NoStop}%
\bibitem [{\citenamefont {Brandelik}\ \emph {et~al.}(1978)\citenamefont
  {Brandelik} \emph {et~al.}}]{DASP:1978dns}%
  \BibitemOpen
  \bibfield  {author} {\bibinfo {author} {\bibfnamefont {R.}~\bibnamefont
  {Brandelik}} \emph {et~al.} (\bibinfo {collaboration} {DASP}),\ }\bibfield
  {title} {\bibinfo {title} {{Total Cross-section for Hadron Production by $e^+
  e^-$ Annihilation at Center-of-mass Energies Between 3.6-{GeV} and
  5.2-{GeV}}},\ }\href {https://doi.org/10.1016/0370-2693(78)90807-9}
  {\bibfield  {journal} {\bibinfo  {journal} {Phys. Lett. B}\ }\textbf
  {\bibinfo {volume} {76}},\ \bibinfo {pages} {361} (\bibinfo {year}
  {1978})}\BibitemShut {NoStop}%
\bibitem [{\citenamefont {Siegrist}\ \emph {et~al.}(1982)\citenamefont
  {Siegrist} \emph {et~al.}}]{Siegrist:1981zp}%
  \BibitemOpen
  \bibfield  {author} {\bibinfo {author} {\bibfnamefont {J.}~\bibnamefont
  {Siegrist}} \emph {et~al.},\ }\bibfield  {title} {\bibinfo {title} {{Hadron
  Production by $e^+ e^-$ Annihilation at Center-Of-Mass Energies Between 2.6
  GeV and 7.8 GeV: Part 1. Total Cross-Section, Multiplicities and Inclusive
  Momentum Distributions}},\ }\href {https://doi.org/10.1103/PhysRevD.26.969}
  {\bibfield  {journal} {\bibinfo  {journal} {Phys. Rev. D}\ }\textbf {\bibinfo
  {volume} {26}},\ \bibinfo {pages} {969} (\bibinfo {year} {1982})}\BibitemShut
  {NoStop}%
\bibitem [{\citenamefont {Bai}\ \emph {et~al.}(2000)\citenamefont {Bai} \emph
  {et~al.}}]{BES:1999wbx}%
  \BibitemOpen
  \bibfield  {author} {\bibinfo {author} {\bibfnamefont {J.~Z.}\ \bibnamefont
  {Bai}} \emph {et~al.} (\bibinfo {collaboration} {BES}),\ }\bibfield  {title}
  {\bibinfo {title} {{Measurement of the total cross-section for hadronic
  production by $e^+ e^-$ annihilation at energies between 2.6 GeV$-$5 GeV}},\
  }\href {https://doi.org/10.1103/PhysRevLett.84.594} {\bibfield  {journal}
  {\bibinfo  {journal} {Phys. Rev. Lett.}\ }\textbf {\bibinfo {volume} {84}},\
  \bibinfo {pages} {594} (\bibinfo {year} {2000})},\ \Eprint
  {https://arxiv.org/abs/hep-ex/9908046} {arXiv:hep-ex/9908046} \BibitemShut
  {NoStop}%
\bibitem [{\citenamefont {Chen}\ \emph
  {et~al.}(2011{\natexlab{a}})\citenamefont {Chen}, \citenamefont {He},\ and\
  \citenamefont {Liu}}]{Chen:2010nv}%
  \BibitemOpen
  \bibfield  {author} {\bibinfo {author} {\bibfnamefont {D.-Y.}\ \bibnamefont
  {Chen}}, \bibinfo {author} {\bibfnamefont {J.}~\bibnamefont {He}},\ and\
  \bibinfo {author} {\bibfnamefont {X.}~\bibnamefont {Liu}},\ }\bibfield
  {title} {\bibinfo {title} {{Nonresonant explanation for the Y(4260) structure
  observed in the $e^+e^-\to J/\psi\pi^+\pi^-$ process}},\ }\href
  {https://doi.org/10.1103/PhysRevD.83.054021} {\bibfield  {journal} {\bibinfo
  {journal} {Phys. Rev. D}\ }\textbf {\bibinfo {volume} {83}},\ \bibinfo
  {pages} {054021} (\bibinfo {year} {2011}{\natexlab{a}})},\ \Eprint
  {https://arxiv.org/abs/1012.5362} {arXiv:1012.5362 [hep-ph]} \BibitemShut
  {NoStop}%
\bibitem [{\citenamefont {Aubert}\ \emph
  {et~al.}(2007{\natexlab{a}})\citenamefont {Aubert} \emph
  {et~al.}}]{BaBar:2006ait}%
  \BibitemOpen
  \bibfield  {author} {\bibinfo {author} {\bibfnamefont {B.}~\bibnamefont
  {Aubert}} \emph {et~al.} (\bibinfo {collaboration} {BaBar}),\ }\bibfield
  {title} {\bibinfo {title} {{Evidence of a broad structure at an invariant
  mass of 4.32 GeV$/c^2$ in the reaction $e^{+} e^{-} \to \pi^{+} \pi^{-}
  \psi(2S)$ measured at BaBar}},\ }\href
  {https://doi.org/10.1103/PhysRevLett.98.212001} {\bibfield  {journal}
  {\bibinfo  {journal} {Phys. Rev. Lett.}\ }\textbf {\bibinfo {volume} {98}},\
  \bibinfo {pages} {212001} (\bibinfo {year} {2007}{\natexlab{a}})},\ \Eprint
  {https://arxiv.org/abs/hep-ex/0610057} {arXiv:hep-ex/0610057} \BibitemShut
  {NoStop}%
\bibitem [{\citenamefont {Chen}\ \emph
  {et~al.}(2011{\natexlab{b}})\citenamefont {Chen}, \citenamefont {He},\ and\
  \citenamefont {Liu}}]{Chen:2011kc}%
  \BibitemOpen
  \bibfield  {author} {\bibinfo {author} {\bibfnamefont {D.-Y.}\ \bibnamefont
  {Chen}}, \bibinfo {author} {\bibfnamefont {J.}~\bibnamefont {He}},\ and\
  \bibinfo {author} {\bibfnamefont {X.}~\bibnamefont {Liu}},\ }\bibfield
  {title} {\bibinfo {title} {{A Novel explanation of charmonium-like structure
  in $e^+e^-\to \psi(2S)\pi^+\pi^-$}},\ }\href
  {https://doi.org/10.1103/PhysRevD.83.074012} {\bibfield  {journal} {\bibinfo
  {journal} {Phys. Rev. D}\ }\textbf {\bibinfo {volume} {83}},\ \bibinfo
  {pages} {074012} (\bibinfo {year} {2011}{\natexlab{b}})},\ \Eprint
  {https://arxiv.org/abs/1101.2474} {arXiv:1101.2474 [hep-ph]} \BibitemShut
  {NoStop}%
\bibitem [{\citenamefont {Chen}\ \emph
  {et~al.}(2016{\natexlab{b}})\citenamefont {Chen}, \citenamefont {Liu},
  \citenamefont {Li},\ and\ \citenamefont {Ke}}]{Chen:2015bft}%
  \BibitemOpen
  \bibfield  {author} {\bibinfo {author} {\bibfnamefont {D.-Y.}\ \bibnamefont
  {Chen}}, \bibinfo {author} {\bibfnamefont {X.}~\bibnamefont {Liu}}, \bibinfo
  {author} {\bibfnamefont {X.-Q.}\ \bibnamefont {Li}},\ and\ \bibinfo {author}
  {\bibfnamefont {H.-W.}\ \bibnamefont {Ke}},\ }\bibfield  {title} {\bibinfo
  {title} {{Unified Fano-like interference picture for charmoniumlike states
  Y(4008), Y(4260) and Y(4360)}},\ }\href
  {https://doi.org/10.1103/PhysRevD.93.014011} {\bibfield  {journal} {\bibinfo
  {journal} {Phys. Rev. D}\ }\textbf {\bibinfo {volume} {93}},\ \bibinfo
  {pages} {014011} (\bibinfo {year} {2016}{\natexlab{b}})},\ \Eprint
  {https://arxiv.org/abs/1512.04157} {arXiv:1512.04157 [hep-ph]} \BibitemShut
  {NoStop}%
\bibitem [{\citenamefont {He}\ \emph {et~al.}(2014)\citenamefont {He},
  \citenamefont {Chen}, \citenamefont {Liu},\ and\ \citenamefont
  {Matsuki}}]{He:2014xna}%
  \BibitemOpen
  \bibfield  {author} {\bibinfo {author} {\bibfnamefont {L.-P.}\ \bibnamefont
  {He}}, \bibinfo {author} {\bibfnamefont {D.-Y.}\ \bibnamefont {Chen}},
  \bibinfo {author} {\bibfnamefont {X.}~\bibnamefont {Liu}},\ and\ \bibinfo
  {author} {\bibfnamefont {T.}~\bibnamefont {Matsuki}},\ }\bibfield  {title}
  {\bibinfo {title} {{Prediction of a missing higher charmonium around 4.26 GeV
  in $J/\psi$ family}},\ }\href
  {https://doi.org/10.1140/epjc/s10052-014-3208-5} {\bibfield  {journal}
  {\bibinfo  {journal} {Eur. Phys. J. C}\ }\textbf {\bibinfo {volume} {74}},\
  \bibinfo {pages} {3208} (\bibinfo {year} {2014})},\ \Eprint
  {https://arxiv.org/abs/1405.3831} {arXiv:1405.3831 [hep-ph]} \BibitemShut
  {NoStop}%
\bibitem [{\citenamefont {Eichten}\ \emph {et~al.}(1980)\citenamefont
  {Eichten}, \citenamefont {Gottfried}, \citenamefont {Kinoshita},
  \citenamefont {Lane},\ and\ \citenamefont {Yan}}]{Eichten:1979ms}%
  \BibitemOpen
  \bibfield  {author} {\bibinfo {author} {\bibfnamefont {E.}~\bibnamefont
  {Eichten}}, \bibinfo {author} {\bibfnamefont {K.}~\bibnamefont {Gottfried}},
  \bibinfo {author} {\bibfnamefont {T.}~\bibnamefont {Kinoshita}}, \bibinfo
  {author} {\bibfnamefont {K.~D.}\ \bibnamefont {Lane}},\ and\ \bibinfo
  {author} {\bibfnamefont {T.-M.}\ \bibnamefont {Yan}},\ }\bibfield  {title}
  {\bibinfo {title} {{Charmonium: Comparison with Experiment}},\ }\href
  {https://doi.org/10.1103/PhysRevD.21.203} {\bibfield  {journal} {\bibinfo
  {journal} {Phys. Rev. D}\ }\textbf {\bibinfo {volume} {21}},\ \bibinfo
  {pages} {203} (\bibinfo {year} {1980})}\BibitemShut {NoStop}%
\bibitem [{\citenamefont {Dong}\ \emph {et~al.}(1994)\citenamefont {Dong},
  \citenamefont {Yu}, \citenamefont {Zhang},\ and\ \citenamefont
  {Shen}}]{Dong:1994zj}%
  \BibitemOpen
  \bibfield  {author} {\bibinfo {author} {\bibfnamefont {Y.-B.}\ \bibnamefont
  {Dong}}, \bibinfo {author} {\bibfnamefont {Y.-W.}\ \bibnamefont {Yu}},
  \bibinfo {author} {\bibfnamefont {Z.-Y.}\ \bibnamefont {Zhang}},\ and\
  \bibinfo {author} {\bibfnamefont {P.-N.}\ \bibnamefont {Shen}},\ }\bibfield
  {title} {\bibinfo {title} {{Leptonic decay of charmonium}},\ }\href
  {https://doi.org/10.1103/PhysRevD.49.1642} {\bibfield  {journal} {\bibinfo
  {journal} {Phys. Rev. D}\ }\textbf {\bibinfo {volume} {49}},\ \bibinfo
  {pages} {1642} (\bibinfo {year} {1994})}\BibitemShut {NoStop}%
\bibitem [{\citenamefont {Ding}\ \emph {et~al.}(1995)\citenamefont {Ding},
  \citenamefont {Chao},\ and\ \citenamefont {Qin}}]{Ding:1995he}%
  \BibitemOpen
  \bibfield  {author} {\bibinfo {author} {\bibfnamefont {Y.-B.}\ \bibnamefont
  {Ding}}, \bibinfo {author} {\bibfnamefont {K.-T.}\ \bibnamefont {Chao}},\
  and\ \bibinfo {author} {\bibfnamefont {D.-H.}\ \bibnamefont {Qin}},\
  }\bibfield  {title} {\bibinfo {title} {{Possible effects of color screening
  and large string tension in heavy quarkonium spectra}},\ }\href
  {https://doi.org/10.1103/PhysRevD.51.5064} {\bibfield  {journal} {\bibinfo
  {journal} {Phys. Rev. D}\ }\textbf {\bibinfo {volume} {51}},\ \bibinfo
  {pages} {5064} (\bibinfo {year} {1995})},\ \Eprint
  {https://arxiv.org/abs/hep-ph/9502409} {arXiv:hep-ph/9502409} \BibitemShut
  {NoStop}%
\bibitem [{\citenamefont {Li}\ and\ \citenamefont {Chao}(2009)}]{Li:2009zu}%
  \BibitemOpen
  \bibfield  {author} {\bibinfo {author} {\bibfnamefont {B.-Q.}\ \bibnamefont
  {Li}}\ and\ \bibinfo {author} {\bibfnamefont {K.-T.}\ \bibnamefont {Chao}},\
  }\bibfield  {title} {\bibinfo {title} {{Higher Charmonia and X,Y,Z states
  with Screened Potential}},\ }\href
  {https://doi.org/10.1103/PhysRevD.79.094004} {\bibfield  {journal} {\bibinfo
  {journal} {Phys. Rev. D}\ }\textbf {\bibinfo {volume} {79}},\ \bibinfo
  {pages} {094004} (\bibinfo {year} {2009})},\ \Eprint
  {https://arxiv.org/abs/0903.5506} {arXiv:0903.5506 [hep-ph]} \BibitemShut
  {NoStop}%
\bibitem [{\citenamefont {Yuan}(2014)}]{Yuan:2013uta}%
  \BibitemOpen
  \bibfield  {author} {\bibinfo {author} {\bibfnamefont {C.-Z.}\ \bibnamefont
  {Yuan}},\ }\bibfield  {title} {\bibinfo {title} {{Evidence for resonant
  structures in $e^{+}e^{-} \to \pi^{+}\pi^{-}h_c$}},\ }\href
  {https://doi.org/10.1088/1674-1137/38/4/043001} {\bibfield  {journal}
  {\bibinfo  {journal} {Chin. Phys. C}\ }\textbf {\bibinfo {volume} {38}},\
  \bibinfo {pages} {043001} (\bibinfo {year} {2014})},\ \Eprint
  {https://arxiv.org/abs/1312.6399} {arXiv:1312.6399 [hep-ex]} \BibitemShut
  {NoStop}%
\bibitem [{\citenamefont {Chen}\ \emph
  {et~al.}(2015{\natexlab{b}})\citenamefont {Chen}, \citenamefont {Liu},\ and\
  \citenamefont {Matsuki}}]{Chen:2014sra}%
  \BibitemOpen
  \bibfield  {author} {\bibinfo {author} {\bibfnamefont {D.-Y.}\ \bibnamefont
  {Chen}}, \bibinfo {author} {\bibfnamefont {X.}~\bibnamefont {Liu}},\ and\
  \bibinfo {author} {\bibfnamefont {T.}~\bibnamefont {Matsuki}},\ }\bibfield
  {title} {\bibinfo {title} {{Observation of $e^+e^-\to \chi_{c0}\omega$ and
  missing higher charmonium $\psi(4S)$}},\ }\href
  {https://doi.org/10.1103/PhysRevD.91.094023} {\bibfield  {journal} {\bibinfo
  {journal} {Phys. Rev. D}\ }\textbf {\bibinfo {volume} {91}},\ \bibinfo
  {pages} {094023} (\bibinfo {year} {2015}{\natexlab{b}})},\ \Eprint
  {https://arxiv.org/abs/1411.5136} {arXiv:1411.5136 [hep-ph]} \BibitemShut
  {NoStop}%
\bibitem [{\citenamefont {Wang}\ \emph {et~al.}(2015)\citenamefont {Wang} \emph
  {et~al.}}]{Belle:2014wyt}%
  \BibitemOpen
  \bibfield  {author} {\bibinfo {author} {\bibfnamefont {X.~L.}\ \bibnamefont
  {Wang}} \emph {et~al.} (\bibinfo {collaboration} {Belle}),\ }\bibfield
  {title} {\bibinfo {title} {{Measurement of $e^+e^- \to \pi^+\pi^-\psi(2S)$
  via Initial State Radiation at Belle}},\ }\href
  {https://doi.org/10.1103/PhysRevD.91.112007} {\bibfield  {journal} {\bibinfo
  {journal} {Phys. Rev. D}\ }\textbf {\bibinfo {volume} {91}},\ \bibinfo
  {pages} {112007} (\bibinfo {year} {2015})},\ \Eprint
  {https://arxiv.org/abs/1410.7641} {arXiv:1410.7641 [hep-ex]} \BibitemShut
  {NoStop}%
\bibitem [{\citenamefont {Ablikim}\ \emph {et~al.}(2013)\citenamefont {Ablikim}
  \emph {et~al.}}]{BESIII:2013ouc}%
  \BibitemOpen
  \bibfield  {author} {\bibinfo {author} {\bibfnamefont {M.}~\bibnamefont
  {Ablikim}} \emph {et~al.} (\bibinfo {collaboration} {BESIII}),\ }\bibfield
  {title} {\bibinfo {title} {{Observation of a Charged Charmoniumlike Structure
  $Z_c$(4020) and Search for the $Z_c$(3900) in $e^+e^- \to \pi^+\pi^-h_c$}},\
  }\href {https://doi.org/10.1103/PhysRevLett.111.242001} {\bibfield  {journal}
  {\bibinfo  {journal} {Phys. Rev. Lett.}\ }\textbf {\bibinfo {volume} {111}},\
  \bibinfo {pages} {242001} (\bibinfo {year} {2013})},\ \Eprint
  {https://arxiv.org/abs/1309.1896} {arXiv:1309.1896 [hep-ex]} \BibitemShut
  {NoStop}%
\bibitem [{\citenamefont {Ablikim}\ \emph
  {et~al.}(2017{\natexlab{a}})\citenamefont {Ablikim} \emph
  {et~al.}}]{BESIII:2016bnd}%
  \BibitemOpen
  \bibfield  {author} {\bibinfo {author} {\bibfnamefont {M.}~\bibnamefont
  {Ablikim}} \emph {et~al.} (\bibinfo {collaboration} {BESIII}),\ }\bibfield
  {title} {\bibinfo {title} {{Precise measurement of the $e^+e^-\to
  \pi^+\pi^-J/\psi$ cross section at center-of-mass energies from 3.77 to 4.60
  GeV}},\ }\href {https://doi.org/10.1103/PhysRevLett.118.092001} {\bibfield
  {journal} {\bibinfo  {journal} {Phys. Rev. Lett.}\ }\textbf {\bibinfo
  {volume} {118}},\ \bibinfo {pages} {092001} (\bibinfo {year}
  {2017}{\natexlab{a}})},\ \Eprint {https://arxiv.org/abs/1611.01317}
  {arXiv:1611.01317 [hep-ex]} \BibitemShut {NoStop}%
\bibitem [{\citenamefont {Ablikim}\ \emph
  {et~al.}(2017{\natexlab{b}})\citenamefont {Ablikim} \emph
  {et~al.}}]{BESIII:2016adj}%
  \BibitemOpen
  \bibfield  {author} {\bibinfo {author} {\bibfnamefont {M.}~\bibnamefont
  {Ablikim}} \emph {et~al.} (\bibinfo {collaboration} {BESIII}),\ }\bibfield
  {title} {\bibinfo {title} {{Evidence of Two Resonant Structures in $e^+ e^-
  \to \pi^+ \pi^- h_c$}},\ }\href
  {https://doi.org/10.1103/PhysRevLett.118.092002} {\bibfield  {journal}
  {\bibinfo  {journal} {Phys. Rev. Lett.}\ }\textbf {\bibinfo {volume} {118}},\
  \bibinfo {pages} {092002} (\bibinfo {year} {2017}{\natexlab{b}})},\ \Eprint
  {https://arxiv.org/abs/1610.07044} {arXiv:1610.07044 [hep-ex]} \BibitemShut
  {NoStop}%
\bibitem [{\citenamefont {Ablikim}\ \emph
  {et~al.}(2017{\natexlab{c}})\citenamefont {Ablikim} \emph
  {et~al.}}]{BESIII:2017tqk}%
  \BibitemOpen
  \bibfield  {author} {\bibinfo {author} {\bibfnamefont {M.}~\bibnamefont
  {Ablikim}} \emph {et~al.} (\bibinfo {collaboration} {BESIII}),\ }\bibfield
  {title} {\bibinfo {title} {{Measurement of $e^{+}e^{-}\rightarrow
  \pi^{+}\pi^{-}\psi(3686)$ from 4.008 to 4.600\textasciitilde{}GeV and
  observation of a charged structure in the $\pi^{\pm}\psi(3686)$ mass
  spectrum}},\ }\href {https://doi.org/10.1103/PhysRevD.96.032004} {\bibfield
  {journal} {\bibinfo  {journal} {Phys. Rev. D}\ }\textbf {\bibinfo {volume}
  {96}},\ \bibinfo {pages} {032004} (\bibinfo {year} {2017}{\natexlab{c}})},\
  \bibinfo {note} {[Erratum: Phys.Rev.D 99, 019903 (2019)]},\ \Eprint
  {https://arxiv.org/abs/1703.08787} {arXiv:1703.08787 [hep-ex]} \BibitemShut
  {NoStop}%
\bibitem [{\citenamefont {Wang}\ \emph {et~al.}(2019)\citenamefont {Wang},
  \citenamefont {Chen}, \citenamefont {Liu},\ and\ \citenamefont
  {Matsuki}}]{Wang:2019mhs}%
  \BibitemOpen
  \bibfield  {author} {\bibinfo {author} {\bibfnamefont {J.-Z.}\ \bibnamefont
  {Wang}}, \bibinfo {author} {\bibfnamefont {D.-Y.}\ \bibnamefont {Chen}},
  \bibinfo {author} {\bibfnamefont {X.}~\bibnamefont {Liu}},\ and\ \bibinfo
  {author} {\bibfnamefont {T.}~\bibnamefont {Matsuki}},\ }\bibfield  {title}
  {\bibinfo {title} {{Constructing $J/\psi$ family with updated data of
  charmoniumlike $Y$ states}},\ }\href
  {https://doi.org/10.1103/PhysRevD.99.114003} {\bibfield  {journal} {\bibinfo
  {journal} {Phys. Rev. D}\ }\textbf {\bibinfo {volume} {99}},\ \bibinfo
  {pages} {114003} (\bibinfo {year} {2019})},\ \Eprint
  {https://arxiv.org/abs/1903.07115} {arXiv:1903.07115 [hep-ph]} \BibitemShut
  {NoStop}%
\bibitem [{\citenamefont {Wang}\ and\ \citenamefont
  {Liu}(2023)}]{Wang:2022jxj}%
  \BibitemOpen
  \bibfield  {author} {\bibinfo {author} {\bibfnamefont {J.-Z.}\ \bibnamefont
  {Wang}}\ and\ \bibinfo {author} {\bibfnamefont {X.}~\bibnamefont {Liu}},\
  }\bibfield  {title} {\bibinfo {title} {{Confirming the existence of a new
  higher charmonium \ensuremath{\psi}(4500) by the newly released data of
  $e^+e^-\to K^+K^-J/\psi$}},\ }\href
  {https://doi.org/10.1103/PhysRevD.107.054016} {\bibfield  {journal} {\bibinfo
   {journal} {Phys. Rev. D}\ }\textbf {\bibinfo {volume} {107}},\ \bibinfo
  {pages} {054016} (\bibinfo {year} {2023})},\ \Eprint
  {https://arxiv.org/abs/2212.13512} {arXiv:2212.13512 [hep-ph]} \BibitemShut
  {NoStop}%
\bibitem [{\citenamefont {Wang}\ and\ \citenamefont
  {Liu}(2024)}]{Wang:2023zxj}%
  \BibitemOpen
  \bibfield  {author} {\bibinfo {author} {\bibfnamefont {J.-Z.}\ \bibnamefont
  {Wang}}\ and\ \bibinfo {author} {\bibfnamefont {X.}~\bibnamefont {Liu}},\
  }\bibfield  {title} {\bibinfo {title} {{Identifying a characterized energy
  level structure of higher charmonium well matched to the peak structures in
  $e^+e^\to\pi^+D^0D^{*-}$}},\ }\href
  {https://doi.org/10.1016/j.physletb.2024.138456} {\bibfield  {journal}
  {\bibinfo  {journal} {Phys. Lett. B}\ }\textbf {\bibinfo {volume} {849}},\
  \bibinfo {pages} {138456} (\bibinfo {year} {2024})},\ \Eprint
  {https://arxiv.org/abs/2306.14695} {arXiv:2306.14695 [hep-ph]} \BibitemShut
  {NoStop}%
\bibitem [{\citenamefont {Peng}\ \emph {et~al.}(2024)\citenamefont {Peng},
  \citenamefont {Bai}, \citenamefont {Wang},\ and\ \citenamefont
  {Liu}}]{Peng:2024xui}%
  \BibitemOpen
  \bibfield  {author} {\bibinfo {author} {\bibfnamefont {T.-C.}\ \bibnamefont
  {Peng}}, \bibinfo {author} {\bibfnamefont {Z.-Y.}\ \bibnamefont {Bai}},
  \bibinfo {author} {\bibfnamefont {J.-Z.}\ \bibnamefont {Wang}},\ and\
  \bibinfo {author} {\bibfnamefont {X.}~\bibnamefont {Liu}},\ }\bibfield
  {title} {\bibinfo {title} {{How higher charmonia shape the puzzling data of
  the e+e\ensuremath{-}\textrightarrow{}\ensuremath{\eta}J/\ensuremath{\psi}
  cross section}},\ }\href {https://doi.org/10.1103/PhysRevD.109.094048}
  {\bibfield  {journal} {\bibinfo  {journal} {Phys. Rev. D}\ }\textbf {\bibinfo
  {volume} {109}},\ \bibinfo {pages} {094048} (\bibinfo {year} {2024})},\
  \Eprint {https://arxiv.org/abs/2403.03705} {arXiv:2403.03705 [hep-ph]}
  \BibitemShut {NoStop}%
\bibitem [{\citenamefont {Ablikim}\ \emph {et~al.}(2019)\citenamefont {Ablikim}
  \emph {et~al.}}]{BESIII:2018iea}%
  \BibitemOpen
  \bibfield  {author} {\bibinfo {author} {\bibfnamefont {M.}~\bibnamefont
  {Ablikim}} \emph {et~al.} (\bibinfo {collaboration} {BESIII}),\ }\bibfield
  {title} {\bibinfo {title} {{Evidence of a resonant structure in the
  $e^+e^-\to \pi^+D^0D^{*-}$ cross section between 4.05 and 4.60 GeV}},\ }\href
  {https://doi.org/10.1103/PhysRevLett.122.102002} {\bibfield  {journal}
  {\bibinfo  {journal} {Phys. Rev. Lett.}\ }\textbf {\bibinfo {volume} {122}},\
  \bibinfo {pages} {102002} (\bibinfo {year} {2019})},\ \Eprint
  {https://arxiv.org/abs/1808.02847} {arXiv:1808.02847 [hep-ex]} \BibitemShut
  {NoStop}%
\bibitem [{\citenamefont {Ablikim}\ \emph {et~al.}(2024)\citenamefont {Ablikim}
  \emph {et~al.}}]{BESIII:2023tll}%
  \BibitemOpen
  \bibfield  {author} {\bibinfo {author} {\bibfnamefont {M.}~\bibnamefont
  {Ablikim}} \emph {et~al.} (\bibinfo {collaboration} {BESIII}),\ }\bibfield
  {title} {\bibinfo {title} {{Measurement of
  e+e-\textrightarrow{}\ensuremath{\eta}J/\ensuremath{\psi} cross section from
  s=3.808\,\,GeV to 4.951~GeV}},\ }\href
  {https://doi.org/10.1103/PhysRevD.109.092012} {\bibfield  {journal} {\bibinfo
   {journal} {Phys. Rev. D}\ }\textbf {\bibinfo {volume} {109}},\ \bibinfo
  {pages} {092012} (\bibinfo {year} {2024})},\ \Eprint
  {https://arxiv.org/abs/2310.03361} {arXiv:2310.03361 [hep-ex]} \BibitemShut
  {NoStop}%
\bibitem [{\citenamefont {Ablikim}\ \emph
  {et~al.}(2022{\natexlab{a}})\citenamefont {Ablikim} \emph
  {et~al.}}]{BESIII:2022joj}%
  \BibitemOpen
  \bibfield  {author} {\bibinfo {author} {\bibfnamefont {M.}~\bibnamefont
  {Ablikim}} \emph {et~al.} (\bibinfo {collaboration} {(BESIII),, BESIII}),\
  }\bibfield  {title} {\bibinfo {title} {{Observation of the Y(4230) and a new
  structure in $e^+e^-\to K^+K^-J/\psi$}},\ }\href
  {https://doi.org/10.1088/1674-1137/ac945c} {\bibfield  {journal} {\bibinfo
  {journal} {Chin. Phys. C}\ }\textbf {\bibinfo {volume} {46}},\ \bibinfo
  {pages} {111002} (\bibinfo {year} {2022}{\natexlab{a}})},\ \Eprint
  {https://arxiv.org/abs/2204.07800} {arXiv:2204.07800 [hep-ex]} \BibitemShut
  {NoStop}%
\bibitem [{\citenamefont {Godfrey}\ and\ \citenamefont
  {Isgur}(1985)}]{Godfrey:1985xj}%
  \BibitemOpen
  \bibfield  {author} {\bibinfo {author} {\bibfnamefont {S.}~\bibnamefont
  {Godfrey}}\ and\ \bibinfo {author} {\bibfnamefont {N.}~\bibnamefont
  {Isgur}},\ }\bibfield  {title} {\bibinfo {title} {{Mesons in a Relativized
  Quark Model with Chromodynamics}},\ }\href
  {https://doi.org/10.1103/PhysRevD.32.189} {\bibfield  {journal} {\bibinfo
  {journal} {Phys. Rev. D}\ }\textbf {\bibinfo {volume} {32}},\ \bibinfo
  {pages} {189} (\bibinfo {year} {1985})}\BibitemShut {NoStop}%
\bibitem [{\citenamefont {Barnes}\ \emph {et~al.}(2005)\citenamefont {Barnes},
  \citenamefont {Godfrey},\ and\ \citenamefont {Swanson}}]{Barnes:2005pb}%
  \BibitemOpen
  \bibfield  {author} {\bibinfo {author} {\bibfnamefont {T.}~\bibnamefont
  {Barnes}}, \bibinfo {author} {\bibfnamefont {S.}~\bibnamefont {Godfrey}},\
  and\ \bibinfo {author} {\bibfnamefont {E.~S.}\ \bibnamefont {Swanson}},\
  }\bibfield  {title} {\bibinfo {title} {{Higher charmonia}},\ }\href
  {https://doi.org/10.1103/PhysRevD.72.054026} {\bibfield  {journal} {\bibinfo
  {journal} {Phys. Rev. D}\ }\textbf {\bibinfo {volume} {72}},\ \bibinfo
  {pages} {054026} (\bibinfo {year} {2005})},\ \Eprint
  {https://arxiv.org/abs/hep-ph/0505002} {arXiv:hep-ph/0505002} \BibitemShut
  {NoStop}%
\bibitem [{\citenamefont {Godfrey}\ and\ \citenamefont
  {Moats}(2016)}]{Godfrey:2015dva}%
  \BibitemOpen
  \bibfield  {author} {\bibinfo {author} {\bibfnamefont {S.}~\bibnamefont
  {Godfrey}}\ and\ \bibinfo {author} {\bibfnamefont {K.}~\bibnamefont
  {Moats}},\ }\bibfield  {title} {\bibinfo {title} {{Properties of Excited
  Charm and Charm-Strange Mesons}},\ }\href
  {https://doi.org/10.1103/PhysRevD.93.034035} {\bibfield  {journal} {\bibinfo
  {journal} {Phys. Rev. D}\ }\textbf {\bibinfo {volume} {93}},\ \bibinfo
  {pages} {034035} (\bibinfo {year} {2016})},\ \Eprint
  {https://arxiv.org/abs/1510.08305} {arXiv:1510.08305 [hep-ph]} \BibitemShut
  {NoStop}%
\bibitem [{\citenamefont {Godfrey}\ and\ \citenamefont
  {Moats}(2015)}]{Godfrey:2015dia}%
  \BibitemOpen
  \bibfield  {author} {\bibinfo {author} {\bibfnamefont {S.}~\bibnamefont
  {Godfrey}}\ and\ \bibinfo {author} {\bibfnamefont {K.}~\bibnamefont
  {Moats}},\ }\bibfield  {title} {\bibinfo {title} {{Bottomonium Mesons and
  Strategies for their Observation}},\ }\href
  {https://doi.org/10.1103/PhysRevD.92.054034} {\bibfield  {journal} {\bibinfo
  {journal} {Phys. Rev. D}\ }\textbf {\bibinfo {volume} {92}},\ \bibinfo
  {pages} {054034} (\bibinfo {year} {2015})},\ \Eprint
  {https://arxiv.org/abs/1507.00024} {arXiv:1507.00024 [hep-ph]} \BibitemShut
  {NoStop}%
\bibitem [{\citenamefont {Godfrey}\ and\ \citenamefont
  {Moats}(2014)}]{Godfrey:2014fga}%
  \BibitemOpen
  \bibfield  {author} {\bibinfo {author} {\bibfnamefont {S.}~\bibnamefont
  {Godfrey}}\ and\ \bibinfo {author} {\bibfnamefont {K.}~\bibnamefont
  {Moats}},\ }\bibfield  {title} {\bibinfo {title} {{The $D_{sJ}^*(2860)$
  Mesons as Excited D-wave $c\bar{s}$ States}},\ }\href
  {https://doi.org/10.1103/PhysRevD.90.117501} {\bibfield  {journal} {\bibinfo
  {journal} {Phys. Rev. D}\ }\textbf {\bibinfo {volume} {90}},\ \bibinfo
  {pages} {117501} (\bibinfo {year} {2014})},\ \Eprint
  {https://arxiv.org/abs/1409.0874} {arXiv:1409.0874 [hep-ph]} \BibitemShut
  {NoStop}%
\bibitem [{\citenamefont {Godfrey}\ \emph {et~al.}(2016)\citenamefont
  {Godfrey}, \citenamefont {Moats},\ and\ \citenamefont
  {Swanson}}]{Godfrey:2016nwn}%
  \BibitemOpen
  \bibfield  {author} {\bibinfo {author} {\bibfnamefont {S.}~\bibnamefont
  {Godfrey}}, \bibinfo {author} {\bibfnamefont {K.}~\bibnamefont {Moats}},\
  and\ \bibinfo {author} {\bibfnamefont {E.~S.}\ \bibnamefont {Swanson}},\
  }\bibfield  {title} {\bibinfo {title} {{$B$ and $B_s$ Meson Spectroscopy}},\
  }\href {https://doi.org/10.1103/PhysRevD.94.054025} {\bibfield  {journal}
  {\bibinfo  {journal} {Phys. Rev. D}\ }\textbf {\bibinfo {volume} {94}},\
  \bibinfo {pages} {054025} (\bibinfo {year} {2016})},\ \Eprint
  {https://arxiv.org/abs/1607.02169} {arXiv:1607.02169 [hep-ph]} \BibitemShut
  {NoStop}%
\bibitem [{\citenamefont {Duan}\ \emph {et~al.}(2020)\citenamefont {Duan},
  \citenamefont {Luo}, \citenamefont {Liu},\ and\ \citenamefont
  {Matsuki}}]{Duan:2020tsx}%
  \BibitemOpen
  \bibfield  {author} {\bibinfo {author} {\bibfnamefont {M.-X.}\ \bibnamefont
  {Duan}}, \bibinfo {author} {\bibfnamefont {S.-Q.}\ \bibnamefont {Luo}},
  \bibinfo {author} {\bibfnamefont {X.}~\bibnamefont {Liu}},\ and\ \bibinfo
  {author} {\bibfnamefont {T.}~\bibnamefont {Matsuki}},\ }\bibfield  {title}
  {\bibinfo {title} {{Possibility of charmoniumlike state $X(3915)$ as
  $\chi_{c0}(2P)$ state}},\ }\href
  {https://doi.org/10.1103/PhysRevD.101.054029} {\bibfield  {journal} {\bibinfo
   {journal} {Phys. Rev. D}\ }\textbf {\bibinfo {volume} {101}},\ \bibinfo
  {pages} {054029} (\bibinfo {year} {2020})},\ \Eprint
  {https://arxiv.org/abs/2002.03311} {arXiv:2002.03311 [hep-ph]} \BibitemShut
  {NoStop}%
\bibitem [{\citenamefont {Rarita}\ and\ \citenamefont
  {Schwinger}(1941)}]{Rarita:1941zza}%
  \BibitemOpen
  \bibfield  {author} {\bibinfo {author} {\bibfnamefont {W.}~\bibnamefont
  {Rarita}}\ and\ \bibinfo {author} {\bibfnamefont {J.}~\bibnamefont
  {Schwinger}},\ }\bibfield  {title} {\bibinfo {title} {{On the Neutron-Proton
  Interaction}},\ }\href {https://doi.org/10.1103/PhysRev.59.436} {\bibfield
  {journal} {\bibinfo  {journal} {Phys. Rev.}\ }\textbf {\bibinfo {volume}
  {59}},\ \bibinfo {pages} {436} (\bibinfo {year} {1941})}\BibitemShut
  {NoStop}%
\bibitem [{\citenamefont {Choi}\ \emph {et~al.}(2003)\citenamefont {Choi} \emph
  {et~al.}}]{Belle:2003nnu}%
  \BibitemOpen
  \bibfield  {author} {\bibinfo {author} {\bibfnamefont {S.~K.}\ \bibnamefont
  {Choi}} \emph {et~al.} (\bibinfo {collaboration} {Belle}),\ }\bibfield
  {title} {\bibinfo {title} {{Observation of a narrow charmonium-like state in
  exclusive $B^\pm \to K^\pm \pi^+ \pi^- J/\psi$ decays}},\ }\href
  {https://doi.org/10.1103/PhysRevLett.91.262001} {\bibfield  {journal}
  {\bibinfo  {journal} {Phys. Rev. Lett.}\ }\textbf {\bibinfo {volume} {91}},\
  \bibinfo {pages} {262001} (\bibinfo {year} {2003})},\ \Eprint
  {https://arxiv.org/abs/hep-ex/0309032} {arXiv:hep-ex/0309032} \BibitemShut
  {NoStop}%
\bibitem [{\citenamefont {Aubert}\ \emph {et~al.}(2003)\citenamefont {Aubert}
  \emph {et~al.}}]{BaBar:2003oey}%
  \BibitemOpen
  \bibfield  {author} {\bibinfo {author} {\bibfnamefont {B.}~\bibnamefont
  {Aubert}} \emph {et~al.} (\bibinfo {collaboration} {BaBar}),\ }\bibfield
  {title} {\bibinfo {title} {{Observation of a narrow meson decaying to $D_s^+
  \pi^0$ at a mass of 2.32 GeV/c$^2$}},\ }\href
  {https://doi.org/10.1103/PhysRevLett.90.242001} {\bibfield  {journal}
  {\bibinfo  {journal} {Phys. Rev. Lett.}\ }\textbf {\bibinfo {volume} {90}},\
  \bibinfo {pages} {242001} (\bibinfo {year} {2003})},\ \Eprint
  {https://arxiv.org/abs/hep-ex/0304021} {arXiv:hep-ex/0304021} \BibitemShut
  {NoStop}%
\bibitem [{\citenamefont {Besson}\ \emph {et~al.}(2003)\citenamefont {Besson}
  \emph {et~al.}}]{CLEO:2003ggt}%
  \BibitemOpen
  \bibfield  {author} {\bibinfo {author} {\bibfnamefont {D.}~\bibnamefont
  {Besson}} \emph {et~al.} (\bibinfo {collaboration} {CLEO}),\ }\bibfield
  {title} {\bibinfo {title} {{Observation of a narrow resonance of mass 2.46
  GeV/c$^2$ decaying to $D_s^{*+}\pi^0$ and confirmation of the
  $D^{*}_{sJ}(2317)$ state}},\ }\href
  {https://doi.org/10.1103/PhysRevD.68.032002} {\bibfield  {journal} {\bibinfo
  {journal} {Phys. Rev. D}\ }\textbf {\bibinfo {volume} {68}},\ \bibinfo
  {pages} {032002} (\bibinfo {year} {2003})},\ \bibinfo {note} {[Erratum:
  Phys.Rev.D 75, 119908 (2007)]},\ \Eprint
  {https://arxiv.org/abs/hep-ex/0305100} {arXiv:hep-ex/0305100} \BibitemShut
  {NoStop}%
\bibitem [{\citenamefont {Aubert}\ \emph
  {et~al.}(2007{\natexlab{b}})\citenamefont {Aubert} \emph
  {et~al.}}]{BaBar:2006itc}%
  \BibitemOpen
  \bibfield  {author} {\bibinfo {author} {\bibfnamefont {B.}~\bibnamefont
  {Aubert}} \emph {et~al.} (\bibinfo {collaboration} {BaBar}),\ }\bibfield
  {title} {\bibinfo {title} {{Observation of a charmed baryon decaying to
  $D^0p$ at a mass near 2.94 GeV/c$^2$}},\ }\href
  {https://doi.org/10.1103/PhysRevLett.98.012001} {\bibfield  {journal}
  {\bibinfo  {journal} {Phys. Rev. Lett.}\ }\textbf {\bibinfo {volume} {98}},\
  \bibinfo {pages} {012001} (\bibinfo {year} {2007}{\natexlab{b}})},\ \Eprint
  {https://arxiv.org/abs/hep-ex/0603052} {arXiv:hep-ex/0603052} \BibitemShut
  {NoStop}%
\bibitem [{\citenamefont {Liu}\ \emph {et~al.}(2025)\citenamefont {Liu},
  \citenamefont {Pan}, \citenamefont {Liu}, \citenamefont {Wu}, \citenamefont
  {Lu},\ and\ \citenamefont {Geng}}]{Liu:2024uxn}%
  \BibitemOpen
  \bibfield  {author} {\bibinfo {author} {\bibfnamefont {M.-Z.}\ \bibnamefont
  {Liu}}, \bibinfo {author} {\bibfnamefont {Y.-W.}\ \bibnamefont {Pan}},
  \bibinfo {author} {\bibfnamefont {Z.-W.}\ \bibnamefont {Liu}}, \bibinfo
  {author} {\bibfnamefont {T.-W.}\ \bibnamefont {Wu}}, \bibinfo {author}
  {\bibfnamefont {J.-X.}\ \bibnamefont {Lu}},\ and\ \bibinfo {author}
  {\bibfnamefont {L.-S.}\ \bibnamefont {Geng}},\ }\bibfield  {title} {\bibinfo
  {title} {{Three ways to decipher the nature of exotic hadrons: Multiplets,
  three-body hadronic molecules, and correlation functions}},\ }\href
  {https://doi.org/10.1016/j.physrep.2024.12.001} {\bibfield  {journal}
  {\bibinfo  {journal} {Phys. Rept.}\ }\textbf {\bibinfo {volume} {1108}},\
  \bibinfo {pages} {1} (\bibinfo {year} {2025})},\ \Eprint
  {https://arxiv.org/abs/2404.06399} {arXiv:2404.06399 [hep-ph]} \BibitemShut
  {NoStop}%
\bibitem [{\citenamefont {van Beveren}\ and\ \citenamefont
  {Rupp}(2003)}]{vanBeveren:2003kd}%
  \BibitemOpen
  \bibfield  {author} {\bibinfo {author} {\bibfnamefont {E.}~\bibnamefont {van
  Beveren}}\ and\ \bibinfo {author} {\bibfnamefont {G.}~\bibnamefont {Rupp}},\
  }\bibfield  {title} {\bibinfo {title} {{Observed $D_s(2317)$ and tentative
  $D(2100\text{--}2300)$ as the charmed cousins of the light scalar nonet}},\
  }\href {https://doi.org/10.1103/PhysRevLett.91.012003} {\bibfield  {journal}
  {\bibinfo  {journal} {Phys. Rev. Lett.}\ }\textbf {\bibinfo {volume} {91}},\
  \bibinfo {pages} {012003} (\bibinfo {year} {2003})},\ \Eprint
  {https://arxiv.org/abs/hep-ph/0305035} {arXiv:hep-ph/0305035} \BibitemShut
  {NoStop}%
\bibitem [{\citenamefont {van Beveren}\ and\ \citenamefont
  {Rupp}(2004)}]{vanBeveren:2003jv}%
  \BibitemOpen
  \bibfield  {author} {\bibinfo {author} {\bibfnamefont {E.}~\bibnamefont {van
  Beveren}}\ and\ \bibinfo {author} {\bibfnamefont {G.}~\bibnamefont {Rupp}},\
  }\bibfield  {title} {\bibinfo {title} {{Continuum bound states $K_L$,
  $D_1(2420)$, $D_{s1}(2536)$ and their partners $K_S$, $D_1(2400)$,
  $D^*_{sJ}(2463)$}},\ }\href {https://doi.org/10.1140/epjc/s2003-01465-0}
  {\bibfield  {journal} {\bibinfo  {journal} {Eur. Phys. J. C}\ }\textbf
  {\bibinfo {volume} {32}},\ \bibinfo {pages} {493} (\bibinfo {year} {2004})},\
  \Eprint {https://arxiv.org/abs/hep-ph/0306051} {arXiv:hep-ph/0306051}
  \BibitemShut {NoStop}%
\bibitem [{\citenamefont {Kalashnikova}(2005)}]{Kalashnikova:2005ui}%
  \BibitemOpen
  \bibfield  {author} {\bibinfo {author} {\bibfnamefont {Y.~S.}\ \bibnamefont
  {Kalashnikova}},\ }\bibfield  {title} {\bibinfo {title} {{Coupled-channel
  model for charmonium levels and an option for X(3872)}},\ }\href
  {https://doi.org/10.1103/PhysRevD.72.034010} {\bibfield  {journal} {\bibinfo
  {journal} {Phys. Rev. D}\ }\textbf {\bibinfo {volume} {72}},\ \bibinfo
  {pages} {034010} (\bibinfo {year} {2005})},\ \Eprint
  {https://arxiv.org/abs/hep-ph/0506270} {arXiv:hep-ph/0506270} \BibitemShut
  {NoStop}%
\bibitem [{\citenamefont {Luo}\ \emph {et~al.}(2020)\citenamefont {Luo},
  \citenamefont {Chen}, \citenamefont {Liu},\ and\ \citenamefont
  {Liu}}]{Luo:2019qkm}%
  \BibitemOpen
  \bibfield  {author} {\bibinfo {author} {\bibfnamefont {S.-Q.}\ \bibnamefont
  {Luo}}, \bibinfo {author} {\bibfnamefont {B.}~\bibnamefont {Chen}}, \bibinfo
  {author} {\bibfnamefont {Z.-W.}\ \bibnamefont {Liu}},\ and\ \bibinfo {author}
  {\bibfnamefont {X.}~\bibnamefont {Liu}},\ }\bibfield  {title} {\bibinfo
  {title} {{Resolving the low mass puzzle of $\Lambda_c(2940)^+$}},\ }\href
  {https://doi.org/10.1140/epjc/s10052-020-7874-1} {\bibfield  {journal}
  {\bibinfo  {journal} {Eur. Phys. J. C}\ }\textbf {\bibinfo {volume} {80}},\
  \bibinfo {pages} {301} (\bibinfo {year} {2020})},\ \Eprint
  {https://arxiv.org/abs/1910.14545} {arXiv:1910.14545 [hep-ph]} \BibitemShut
  {NoStop}%
\bibitem [{\citenamefont {Man}\ \emph {et~al.}(2024)\citenamefont {Man},
  \citenamefont {Shu}, \citenamefont {Liu},\ and\ \citenamefont
  {Chen}}]{Man:2024mvl}%
  \BibitemOpen
  \bibfield  {author} {\bibinfo {author} {\bibfnamefont {Z.-L.}\ \bibnamefont
  {Man}}, \bibinfo {author} {\bibfnamefont {C.-R.}\ \bibnamefont {Shu}},
  \bibinfo {author} {\bibfnamefont {Y.-R.}\ \bibnamefont {Liu}},\ and\ \bibinfo
  {author} {\bibfnamefont {H.}~\bibnamefont {Chen}},\ }\bibfield  {title}
  {\bibinfo {title} {{Charmonium states in a coupled-channel model}},\ }\href
  {https://doi.org/10.1140/epjc/s10052-024-13132-7} {\bibfield  {journal}
  {\bibinfo  {journal} {Eur. Phys. J. C}\ }\textbf {\bibinfo {volume} {84}},\
  \bibinfo {pages} {810} (\bibinfo {year} {2024})},\ \Eprint
  {https://arxiv.org/abs/2402.02765} {arXiv:2402.02765 [hep-ph]} \BibitemShut
  {NoStop}%
\bibitem [{\citenamefont {Wang}\ \emph {et~al.}()\citenamefont {Wang},
  \citenamefont {Sun}, \citenamefont {Liu},\ and\ \citenamefont
  {Matsuki}}]{Wang:2018rjg}%
  \BibitemOpen
  \bibfield  {author} {\bibinfo {author} {\bibfnamefont {J.-Z.}\ \bibnamefont
  {Wang}}, \bibinfo {author} {\bibfnamefont {Z.-F.}\ \bibnamefont {Sun}},
  \bibinfo {author} {\bibfnamefont {X.}~\bibnamefont {Liu}},\ and\ \bibinfo
  {author} {\bibfnamefont {T.}~\bibnamefont {Matsuki}},\ }\bibfield  {title}
  {\bibinfo {title} {{Higher bottomonium zoo}},\ }\href
  {https://doi.org/10.1140/epjc/s10052-018-6372-1} {\bibfield  {journal}
  {\bibinfo  {journal} {Eur. Phys. J. C}\ }\textbf {\bibinfo {volume} {78}},\
  \bibinfo {pages} {915}},\ \Eprint {https://arxiv.org/abs/1802.04938}
  {arXiv:1802.04938 [hep-ph]} \BibitemShut {NoStop}%
\bibitem [{\citenamefont {Pan}\ \emph {et~al.}(2024)\citenamefont {Pan},
  \citenamefont {Liu}, \citenamefont {Man},\ and\ \citenamefont
  {Liu}}]{Pan:2024xec}%
  \BibitemOpen
  \bibfield  {author} {\bibinfo {author} {\bibfnamefont {Z.-H.}\ \bibnamefont
  {Pan}}, \bibinfo {author} {\bibfnamefont {C.-X.}\ \bibnamefont {Liu}},
  \bibinfo {author} {\bibfnamefont {Z.-L.}\ \bibnamefont {Man}},\ and\ \bibinfo
  {author} {\bibfnamefont {X.}~\bibnamefont {Liu}},\ }\bibfield  {title}
  {\bibinfo {title} {{Prospects for observing $2D$ and $1F$ charmonium states
  near 4 GeV}},\ }\href@noop {} {\  (\bibinfo {year} {2024})},\ \Eprint
  {https://arxiv.org/abs/2411.15689} {arXiv:2411.15689 [hep-ph]} \BibitemShut
  {NoStop}%
\bibitem [{\citenamefont {Wang}\ \emph {et~al.}(2020)\citenamefont {Wang},
  \citenamefont {Qian}, \citenamefont {Liu},\ and\ \citenamefont
  {Matsuki}}]{Wang:2020prx}%
  \BibitemOpen
  \bibfield  {author} {\bibinfo {author} {\bibfnamefont {J.-Z.}\ \bibnamefont
  {Wang}}, \bibinfo {author} {\bibfnamefont {R.-Q.}\ \bibnamefont {Qian}},
  \bibinfo {author} {\bibfnamefont {X.}~\bibnamefont {Liu}},\ and\ \bibinfo
  {author} {\bibfnamefont {T.}~\bibnamefont {Matsuki}},\ }\bibfield  {title}
  {\bibinfo {title} {{Are the $Y$ states around 4.6 GeV from $e^+e^-$
  annihilation higher charmonia?}},\ }\href
  {https://doi.org/10.1103/PhysRevD.101.034001} {\bibfield  {journal} {\bibinfo
   {journal} {Phys. Rev. D}\ }\textbf {\bibinfo {volume} {101}},\ \bibinfo
  {pages} {034001} (\bibinfo {year} {2020})},\ \Eprint
  {https://arxiv.org/abs/2001.00175} {arXiv:2001.00175 [hep-ph]} \BibitemShut
  {NoStop}%
\bibitem [{\citenamefont {Lu}\ \emph {et~al.}(2016)\citenamefont {Lu},
  \citenamefont {Anwar},\ and\ \citenamefont {Zou}}]{Lu:2016mbb}%
  \BibitemOpen
  \bibfield  {author} {\bibinfo {author} {\bibfnamefont {Y.}~\bibnamefont
  {Lu}}, \bibinfo {author} {\bibfnamefont {M.~N.}\ \bibnamefont {Anwar}},\ and\
  \bibinfo {author} {\bibfnamefont {B.-S.}\ \bibnamefont {Zou}},\ }\bibfield
  {title} {\bibinfo {title} {{Coupled-Channel Effects for the Bottomonium with
  Realistic Wave Functions}},\ }\href
  {https://doi.org/10.1103/PhysRevD.94.034021} {\bibfield  {journal} {\bibinfo
  {journal} {Phys. Rev. D}\ }\textbf {\bibinfo {volume} {94}},\ \bibinfo
  {pages} {034021} (\bibinfo {year} {2016})},\ \Eprint
  {https://arxiv.org/abs/1606.06927} {arXiv:1606.06927 [hep-ph]} \BibitemShut
  {NoStop}%
\bibitem [{\citenamefont {Fu}\ and\ \citenamefont {Jiang}(2019)}]{Fu:2018yxq}%
  \BibitemOpen
  \bibfield  {author} {\bibinfo {author} {\bibfnamefont {H.-F.}\ \bibnamefont
  {Fu}}\ and\ \bibinfo {author} {\bibfnamefont {L.}~\bibnamefont {Jiang}},\
  }\bibfield  {title} {\bibinfo {title} {{Coupled-channel-induced $S{-}D$
  mixing of Charmonia and testing possible assignments for $Y$(4260) and
  $Y$(4360)}},\ }\href {https://doi.org/10.1140/epjc/s10052-019-6976-0}
  {\bibfield  {journal} {\bibinfo  {journal} {Eur. Phys. J. C}\ }\textbf
  {\bibinfo {volume} {79}},\ \bibinfo {pages} {460} (\bibinfo {year} {2019})},\
  \Eprint {https://arxiv.org/abs/1812.00179} {arXiv:1812.00179 [hep-ph]}
  \BibitemShut {NoStop}%
\bibitem [{\citenamefont {Li}\ \emph {et~al.}(2009)\citenamefont {Li},
  \citenamefont {Meng},\ and\ \citenamefont {Chao}}]{Li:2009ad}%
  \BibitemOpen
  \bibfield  {author} {\bibinfo {author} {\bibfnamefont {B.-Q.}\ \bibnamefont
  {Li}}, \bibinfo {author} {\bibfnamefont {C.}~\bibnamefont {Meng}},\ and\
  \bibinfo {author} {\bibfnamefont {K.-T.}\ \bibnamefont {Chao}},\ }\bibfield
  {title} {\bibinfo {title} {{Coupled-Channel and Screening Effects in
  Charmonium Spectrum}},\ }\href {https://doi.org/10.1103/PhysRevD.80.014012}
  {\bibfield  {journal} {\bibinfo  {journal} {Phys. Rev. D}\ }\textbf {\bibinfo
  {volume} {80}},\ \bibinfo {pages} {014012} (\bibinfo {year} {2009})},\
  \Eprint {https://arxiv.org/abs/0904.4068} {arXiv:0904.4068 [hep-ph]}
  \BibitemShut {NoStop}%
\bibitem [{\citenamefont {Duan}\ and\ \citenamefont
  {Liu}(2021)}]{Duan:2021alw}%
  \BibitemOpen
  \bibfield  {author} {\bibinfo {author} {\bibfnamefont {M.-X.}\ \bibnamefont
  {Duan}}\ and\ \bibinfo {author} {\bibfnamefont {X.}~\bibnamefont {Liu}},\
  }\bibfield  {title} {\bibinfo {title} {{Where are 3P and higher P-wave states
  in the charmonium family?}},\ }\href
  {https://doi.org/10.1103/PhysRevD.104.074010} {\bibfield  {journal} {\bibinfo
   {journal} {Phys. Rev. D}\ }\textbf {\bibinfo {volume} {104}},\ \bibinfo
  {pages} {074010} (\bibinfo {year} {2021})},\ \Eprint
  {https://arxiv.org/abs/2107.14438} {arXiv:2107.14438 [hep-ph]} \BibitemShut
  {NoStop}%
\bibitem [{\citenamefont {Pennington}\ and\ \citenamefont
  {Wilson}(2007)}]{Pennington:2007xr}%
  \BibitemOpen
  \bibfield  {author} {\bibinfo {author} {\bibfnamefont {M.~R.}\ \bibnamefont
  {Pennington}}\ and\ \bibinfo {author} {\bibfnamefont {D.~J.}\ \bibnamefont
  {Wilson}},\ }\bibfield  {title} {\bibinfo {title} {{Decay channels and
  charmonium mass-shifts}},\ }\href
  {https://doi.org/10.1103/PhysRevD.76.077502} {\bibfield  {journal} {\bibinfo
  {journal} {Phys. Rev. D}\ }\textbf {\bibinfo {volume} {76}},\ \bibinfo
  {pages} {077502} (\bibinfo {year} {2007})},\ \Eprint
  {https://arxiv.org/abs/0704.3384} {arXiv:0704.3384 [hep-ph]} \BibitemShut
  {NoStop}%
\bibitem [{\citenamefont {Ni}\ \emph {et~al.}(2024)\citenamefont {Ni},
  \citenamefont {Wu},\ and\ \citenamefont {Zhong}}]{Ni:2023lvx}%
  \BibitemOpen
  \bibfield  {author} {\bibinfo {author} {\bibfnamefont {R.-H.}\ \bibnamefont
  {Ni}}, \bibinfo {author} {\bibfnamefont {J.-J.}\ \bibnamefont {Wu}},\ and\
  \bibinfo {author} {\bibfnamefont {X.-H.}\ \bibnamefont {Zhong}},\ }\bibfield
  {title} {\bibinfo {title} {{Unified unquenched quark model for heavy-light
  mesons with chiral dynamics}},\ }\href
  {https://doi.org/10.1103/PhysRevD.109.116006} {\bibfield  {journal} {\bibinfo
   {journal} {Phys. Rev. D}\ }\textbf {\bibinfo {volume} {109}},\ \bibinfo
  {pages} {116006} (\bibinfo {year} {2024})},\ \Eprint
  {https://arxiv.org/abs/2312.04765} {arXiv:2312.04765 [hep-ph]} \BibitemShut
  {NoStop}%
\bibitem [{\citenamefont {Deng}\ \emph {et~al.}(2024)\citenamefont {Deng},
  \citenamefont {Ni}, \citenamefont {Li},\ and\ \citenamefont
  {Zhong}}]{Deng:2023mza}%
  \BibitemOpen
  \bibfield  {author} {\bibinfo {author} {\bibfnamefont {Q.}~\bibnamefont
  {Deng}}, \bibinfo {author} {\bibfnamefont {R.-H.}\ \bibnamefont {Ni}},
  \bibinfo {author} {\bibfnamefont {Q.}~\bibnamefont {Li}},\ and\ \bibinfo
  {author} {\bibfnamefont {X.-H.}\ \bibnamefont {Zhong}},\ }\bibfield  {title}
  {\bibinfo {title} {{Charmonia in an unquenched quark model}},\ }\href
  {https://doi.org/10.1103/PhysRevD.110.056034} {\bibfield  {journal} {\bibinfo
   {journal} {Phys. Rev. D}\ }\textbf {\bibinfo {volume} {110}},\ \bibinfo
  {pages} {056034} (\bibinfo {year} {2024})},\ \Eprint
  {https://arxiv.org/abs/2312.10296} {arXiv:2312.10296 [hep-ph]} \BibitemShut
  {NoStop}%
\bibitem [{\citenamefont {Le~Yaouanc}\ \emph {et~al.}(1973)\citenamefont
  {Le~Yaouanc}, \citenamefont {Oliver}, \citenamefont {Pene},\ and\
  \citenamefont {Raynal}}]{LeYaouanc:1972vsx}%
  \BibitemOpen
  \bibfield  {author} {\bibinfo {author} {\bibfnamefont {A.}~\bibnamefont
  {Le~Yaouanc}}, \bibinfo {author} {\bibfnamefont {L.}~\bibnamefont {Oliver}},
  \bibinfo {author} {\bibfnamefont {O.}~\bibnamefont {Pene}},\ and\ \bibinfo
  {author} {\bibfnamefont {J.~C.}\ \bibnamefont {Raynal}},\ }\bibfield  {title}
  {\bibinfo {title} {{Naive quark pair creation model of strong interaction
  vertices}},\ }\href {https://doi.org/10.1103/PhysRevD.8.2223} {\bibfield
  {journal} {\bibinfo  {journal} {Phys. Rev. D}\ }\textbf {\bibinfo {volume}
  {8}},\ \bibinfo {pages} {2223} (\bibinfo {year} {1973})}\BibitemShut
  {NoStop}%
\bibitem [{\citenamefont {Le~Yaouanc}\ \emph {et~al.}(1974)\citenamefont
  {Le~Yaouanc}, \citenamefont {Oliver}, \citenamefont {Pene},\ and\
  \citenamefont {Raynal}}]{LeYaouanc:1973ldf}%
  \BibitemOpen
  \bibfield  {author} {\bibinfo {author} {\bibfnamefont {A.}~\bibnamefont
  {Le~Yaouanc}}, \bibinfo {author} {\bibfnamefont {L.}~\bibnamefont {Oliver}},
  \bibinfo {author} {\bibfnamefont {O.}~\bibnamefont {Pene}},\ and\ \bibinfo
  {author} {\bibfnamefont {J.~C.}\ \bibnamefont {Raynal}},\ }\bibfield  {title}
  {\bibinfo {title} {{Naive quark pair creation model and baryon decays}},\
  }\href {https://doi.org/10.1103/PhysRevD.9.1415} {\bibfield  {journal}
  {\bibinfo  {journal} {Phys. Rev. D}\ }\textbf {\bibinfo {volume} {9}},\
  \bibinfo {pages} {1415} (\bibinfo {year} {1974})}\BibitemShut {NoStop}%
\bibitem [{\citenamefont {Navas}\ \emph {et~al.}(2024)\citenamefont {Navas}
  \emph {et~al.}}]{ParticleDataGroup:2024cfk}%
  \BibitemOpen
  \bibfield  {author} {\bibinfo {author} {\bibfnamefont {S.}~\bibnamefont
  {Navas}} \emph {et~al.} (\bibinfo {collaboration} {Particle Data Group}),\
  }\bibfield  {title} {\bibinfo {title} {{Review of particle physics}},\ }\href
  {https://doi.org/10.1103/PhysRevD.110.030001} {\bibfield  {journal} {\bibinfo
   {journal} {Phys. Rev. D}\ }\textbf {\bibinfo {volume} {110}},\ \bibinfo
  {pages} {030001} (\bibinfo {year} {2024})}\BibitemShut {NoStop}%
\bibitem [{\citenamefont {Cahn}\ and\ \citenamefont
  {Jackson}(2003)}]{Cahn:2003cw}%
  \BibitemOpen
  \bibfield  {author} {\bibinfo {author} {\bibfnamefont {R.~N.}\ \bibnamefont
  {Cahn}}\ and\ \bibinfo {author} {\bibfnamefont {J.~D.}\ \bibnamefont
  {Jackson}},\ }\bibfield  {title} {\bibinfo {title} {{Spin orbit and tensor
  forces in heavy quark light quark mesons: Implications of the new $D_s$ state
  at 2.32 GeV}},\ }\href {https://doi.org/10.1103/PhysRevD.68.037502}
  {\bibfield  {journal} {\bibinfo  {journal} {Phys. Rev. D}\ }\textbf {\bibinfo
  {volume} {68}},\ \bibinfo {pages} {037502} (\bibinfo {year} {2003})},\
  \Eprint {https://arxiv.org/abs/hep-ph/0305012} {arXiv:hep-ph/0305012}
  \BibitemShut {NoStop}%
\bibitem [{\citenamefont {Li}\ and\ \citenamefont
  {Voloshin}(2013)}]{Li:2013yka}%
  \BibitemOpen
  \bibfield  {author} {\bibinfo {author} {\bibfnamefont {X.}~\bibnamefont
  {Li}}\ and\ \bibinfo {author} {\bibfnamefont {M.~B.}\ \bibnamefont
  {Voloshin}},\ }\bibfield  {title} {\bibinfo {title} {{Suppression of the
  $S$-wave production of $(3/2)^+$ + $(1/2)^-$ heavy meson pairs in $e^+e^-$
  annihilation}},\ }\href {https://doi.org/10.1103/PhysRevD.88.034012}
  {\bibfield  {journal} {\bibinfo  {journal} {Phys. Rev. D}\ }\textbf {\bibinfo
  {volume} {88}},\ \bibinfo {pages} {034012} (\bibinfo {year} {2013})},\
  \Eprint {https://arxiv.org/abs/1307.1072} {arXiv:1307.1072 [hep-ph]}
  \BibitemShut {NoStop}%
\bibitem [{\citenamefont {Ablikim}\ \emph
  {et~al.}(2021{\natexlab{a}})\citenamefont {Ablikim} \emph
  {et~al.}}]{BESIII:2021yal}%
  \BibitemOpen
  \bibfield  {author} {\bibinfo {author} {\bibfnamefont {M.}~\bibnamefont
  {Ablikim}} \emph {et~al.} (\bibinfo {collaboration} {BESIII}),\ }\bibfield
  {title} {\bibinfo {title} {{Measurement of $e^+e^-\to\gamma \chi_{c0,c1,c2}$
  cross sections at center-of-mass energies between 3.77 and 4.60~GeV}},\
  }\href {https://doi.org/10.1103/PhysRevD.104.092001} {\bibfield  {journal}
  {\bibinfo  {journal} {Phys. Rev. D}\ }\textbf {\bibinfo {volume} {104}},\
  \bibinfo {pages} {092001} (\bibinfo {year} {2021}{\natexlab{a}})},\ \Eprint
  {https://arxiv.org/abs/2107.03604} {arXiv:2107.03604 [hep-ex]} \BibitemShut
  {NoStop}%
\bibitem [{\citenamefont {Ablikim}\ \emph
  {et~al.}(2021{\natexlab{b}})\citenamefont {Ablikim} \emph
  {et~al.}}]{BESIII:2021njb}%
  \BibitemOpen
  \bibfield  {author} {\bibinfo {author} {\bibfnamefont {M.}~\bibnamefont
  {Ablikim}} \emph {et~al.} (\bibinfo {collaboration} {BESIII}),\ }\bibfield
  {title} {\bibinfo {title} {{Cross section measurement of
  $e^+e^-\rightarrow\pi^+\pi^-(3686)$ from $\sqrt{S}=4.0076$ to 4.6984~GeV}},\
  }\href {https://doi.org/10.1103/PhysRevD.104.052012} {\bibfield  {journal}
  {\bibinfo  {journal} {Phys. Rev. D}\ }\textbf {\bibinfo {volume} {104}},\
  \bibinfo {pages} {052012} (\bibinfo {year} {2021}{\natexlab{b}})},\ \Eprint
  {https://arxiv.org/abs/2107.09210} {arXiv:2107.09210 [hep-ex]} \BibitemShut
  {NoStop}%
\bibitem [{\citenamefont {Ablikim}\ \emph
  {et~al.}(2022{\natexlab{b}})\citenamefont {Ablikim} \emph
  {et~al.}}]{BESIII:2022qal}%
  \BibitemOpen
  \bibfield  {author} {\bibinfo {author} {\bibfnamefont {M.}~\bibnamefont
  {Ablikim}} \emph {et~al.} (\bibinfo {collaboration} {BESIII}),\ }\bibfield
  {title} {\bibinfo {title} {{Study of the resonance structures in the process
  $e^+e^+ \to \pi^+\pi^- J/ \psi$}},\ }\href
  {https://doi.org/10.1103/PhysRevD.106.072001} {\bibfield  {journal} {\bibinfo
   {journal} {Phys. Rev. D}\ }\textbf {\bibinfo {volume} {106}},\ \bibinfo
  {pages} {072001} (\bibinfo {year} {2022}{\natexlab{b}})},\ \Eprint
  {https://arxiv.org/abs/2206.08554} {arXiv:2206.08554 [hep-ex]} \BibitemShut
  {NoStop}%
\bibitem [{\citenamefont {Ablikim}\ \emph
  {et~al.}(2022{\natexlab{c}})\citenamefont {Ablikim} \emph
  {et~al.}}]{BESIII:2022quc}%
  \BibitemOpen
  \bibfield  {author} {\bibinfo {author} {\bibfnamefont {M.}~\bibnamefont
  {Ablikim}} \emph {et~al.} (\bibinfo {collaboration} {BESIII}),\ }\bibfield
  {title} {\bibinfo {title} {{Measurement of
  $e^{+}e^{-}\rightarrow\pi^{+}\pi^{-}D^{+}D^{-}$ cross sections at
  center-of-mass energies from 4.190 to 4.946 GeV}},\ }\href
  {https://doi.org/10.1103/PhysRevD.106.052012} {\bibfield  {journal} {\bibinfo
   {journal} {Phys. Rev. D}\ }\textbf {\bibinfo {volume} {106}},\ \bibinfo
  {pages} {052012} (\bibinfo {year} {2022}{\natexlab{c}})},\ \Eprint
  {https://arxiv.org/abs/2208.00099} {arXiv:2208.00099 [hep-ex]} \BibitemShut
  {NoStop}%
\bibitem [{\citenamefont {Ni}\ \emph {et~al.}(2025)\citenamefont {Ni},
  \citenamefont {Deng}, \citenamefont {Wu},\ and\ \citenamefont
  {Zhong}}]{Ni:2025gvx}%
  \BibitemOpen
  \bibfield  {author} {\bibinfo {author} {\bibfnamefont {R.-H.}\ \bibnamefont
  {Ni}}, \bibinfo {author} {\bibfnamefont {Q.}~\bibnamefont {Deng}}, \bibinfo
  {author} {\bibfnamefont {J.-J.}\ \bibnamefont {Wu}},\ and\ \bibinfo {author}
  {\bibfnamefont {X.-H.}\ \bibnamefont {Zhong}},\ }\bibfield  {title} {\bibinfo
  {title} {{Bottomonia in an unquenched quark model}},\ }\href@noop {} {\
  (\bibinfo {year} {2025})},\ \Eprint {https://arxiv.org/abs/2501.15110}
  {arXiv:2501.15110 [hep-ph]} \BibitemShut {NoStop}%
\end{thebibliography}%
\end{document}